\documentclass[referee]{aa}

\usepackage{graphicx}
\usepackage{txfonts}

\bibpunct{(}{)}{;}{a}{}{,} 

\usepackage{subfig}
\usepackage{float}
\usepackage{comment}
\usepackage{algorithm,algorithmicx,algpseudocode}
\usepackage{amsmath}
\usepackage{bm}
\usepackage{amsfonts}
\usepackage{booktabs}
\usepackage{amssymb}
\usepackage{array}
\usepackage{url}
\usepackage{epstopdf}
\usepackage{color}
\usepackage{multirow}
\usepackage{natbib}

\newcommand{\R}{\mathbb{R}}

\begin{document}

\title{A count-based imaging model for the {\em{Spectrometer/Telescope for Imaging X-rays (STIX)}} in {\em{Solar Orbiter}}}

\author{Paolo Massa\inst{1}  \and Michele Piana\inst{1,2} \and Anna Maria Massone\inst{1,2} \and Federico Benvenuto \inst{1}}

\institute{Dipartimento di Matematica,  Universit\`{a} degli Studi di Genova, Via   Dodecaneso 35, 16146 Genova, Italy
\and
CNR-SPIN Genova, Via Dodecaneso 33, 16146 Genova, Italy}


\abstract{
The {\em{Spectrometer/Telescope for Imaging X-rays (STIX)}} will look at solar flares across the hard X-ray window provided by the {\em{Solar Orbiter}} cluster. Similarly to the {\em{Reuven Ramaty High Energy Solar Spectroscopic Imager (RHESSI)}}, {\em{STIX}} is a visibility-based imaging instrument, which will ask for Fourier-based image reconstruction methods. However, in this paper we show that, as for {\em{RHESSI}}, also for {\em{STIX}} count-based imaging is possible. Specifically, here we introduce and illustrate a mathematical model that mimics the {\em{STIX}} data formation process as a projection from the incoming photon flux into a vector made of $120$ count components. Then we test the reliability of Expectation Maximization for image reconstruction in the case of several simulated configurations typical of flare morphology.}


\titlerunning{{\em{STIX}} imaging model}
\authorrunning{P. Massa}

\maketitle
	
	\section{Introduction}
The {\em{Spectrometer/Telescope for Imaging X-rays (STIX)}} \citep{2012SPIE.8443E..3LB} will be launched by the {\em{European Space Agency (ESA)}} as part of the {\em{Solar Orbiter}} \citep{2013SoPh..285...25M} payload in order to provide a window on hard X-ray radiation emitted during solar flares in the energy range between a few and a few hundreds keV. The main scientific goal of this instrument is to allow the determination of timing, location, and spectrum of accelerated electrons by measuring timing, location, and spectrum of their photon signature \citep{1992SoPh..137..121J,2003ApJ...595L..97H,2003ApJ...595L.127P,2006ApJ...643..523B,2007ApJ...665..846P,2011SSRv..159..301K,2012ApJ...755...32G,2012A&A...543A..53G,2012ApJ...751..129T,2013ApJ...766...28G,2016ApJ...831..119H,2018ApJ...867...82D,2018A&A...612A..64S}. Specifically, {\em{STIX}} will convey the hard X-ray radiation from the Sun through $30$ pairs of tungsten grids mounted at the extremity of an aluminum tube and correspondingly record the signal using $30$ Cadmium-Telluride detectors made of four rectangular pixels and mounted behind each grid pair \citep{2013SPIE.8903E..1VP}. A rigorous mathematical modeling of this signal formation process \citep{gietal15} based on some numerical approximations, showed that each system, made of two grids and a detector made of four pixels, provides a spatial Fourier component of the incoming flux, named {\em{visibility}}. Therefore, as in radio-interferometry and in the case of the {\em{Reuven Ramaty High Energy Solar Spectroscopic Imager (RHESSI)}} \citep{2002SoPh..210....3L}, also for the {\em{STIX}} image reconstruction problem will be the inverse problem of inverting the Fourier transform of the photon flux from the limited data represented by complex visibilities sampled in the spatial frequency domain, named the $(u,v)$ plane \citep{2002SoPh..210...61H,2009ApJ...703.2004M,2002SoPh..210..193A,2018A&A...615A..59D,2017ApJ...849...10F,2018ApJ...862...68S}. Such an imaging approach has some unquestionable advantages: for example, it may rely on a vast computational corpus of techniques developed in many different domains; also, this kind of algorithms may exploit Fast Fourier Transform to obtain reconstructions within a low computational burden. On the other hand, in the specific case of {\em{STIX}}, the sampling of the $(u,v)$ plane is very sparse (just $30$ visibilities are provided by the telescope, to compare to the more than one hundred visibilities provided by a typical {\em{RHESS}} observation) and therefore any imaging method has to address a significant effort in order to reduce the artifacts possibly induced by the under-sampling of the frequency plane. Further, as shown by \citet{gietal15}, {\em{STIX}} visibilities are complex numbers whose real and imaginary parts are obtained by differences between count measurements: since the latter are Poisson data, visibility statistic is characterized by a lower signal-to-noise ratio than the count one. 

The present paper contains two main results. The first one is a mathematical model showing that count-based imaging is possible in the case of {\em{STIX}} data. Specifically, the model builds up the matrix reproducing the projection of the photon flux onto the count measurements recorded in each {\em{STIX}} pixel. The second result shows that a rather standard count-based imaging method, namely Expectation Maximization \citep{1974AJ.....79..745L,2013A&A...555A..61B}, is able to reconstruct photon flux images with a reliability comparable to the one provided by standard visibility-based methods (and to some extent even higher). The tests made in the paper utilize synthetic data corresponding to physically plausible configurations of flaring events and produced by using the up-to-date version of the {\em{STIX}} Data Processing Software (DPS).  

The paper is organized as follows. Section 2 introduces the count-based imaging model. Section 3 briefly overviews the count-based image reconstruction method. Section 4 shows the results of the image reconstruction procedure. Our conclusions are offered in Section 5.

\section{Count-based imaging model}
{\em{STIX}} consists of thirty pairs of grids, called sub-collimators, that modulate the photon flux incident onto the telescope; in each sub-collimator, the rear and the front grid have different orientations and pitches. Behind each sub-collimator there is a Cadmium-Telluride detector which is vertically divided into four identical pixels. Photons passing through the grids produce typical \textit{Moiré patterns} \citep{2013SPIE.8903E..1VP,gietal15} on the detectors, and are independently recorded by the pixels. Raw data measured by \textit{STIX} are therefore $120$ sets of photon counts. More precisely, it is shown in \citet{gietal15} that the number of counts $A_j, B_j, C_j,D_j$ recorded by the four pixels in the $j$-th detector is approximated by
\begin{equation}\label{count1-a}
A_j \simeq M_0 V({\mathbf{0}}) - M_1 \exp\left(i\frac{\pi}{4}\right) V(-{\boldsymbol{\xi}}_j) - M_1 \exp\left(-i\frac{\pi}{4}\right) V({\boldsymbol{\xi}}_j),
\end{equation}
\begin{equation}\label{count1-b}
B_j \simeq M_0 V({\mathbf{0}}) - i M_1 \exp\left(i\frac{\pi}{4}\right) V(-{\boldsymbol{\xi}}_j) + i M_1 \exp\left(-i\frac{\pi}{4}\right) V({\boldsymbol{\xi}}_j),
\end{equation}
\begin{equation}\label{count1-c}
C_j \simeq M_0 V({\mathbf{0}}) + M_1 \exp\left(i\frac{\pi}{4}\right) V(-{\boldsymbol{\xi}}_j) + M_1 \exp\left(-i\frac{\pi}{4}\right) V({\boldsymbol{\xi}}_j),
\end{equation}
and
\begin{equation}\label{count1-d}
D_j \simeq M_0 V({\mathbf{0}}) + i M_1 \exp\left(i\frac{\pi}{4}\right) V(-{\boldsymbol{\xi}}_j) - i M_1 \exp\left(-i\frac{\pi}{4}\right) V({\boldsymbol{\xi}}_j).
\end{equation}
In these equations we have that
\begin{equation}\label{emme}
M_0 = \frac{l h}{4}~~~~~M_1 = \frac{4}{\pi^3} lh \sin\left(\frac{\pi}{4}\right)
\end{equation}
\begin{equation}\label{csi}
{\boldsymbol{\xi}}_j = {\mathbf{k}}_j^f \frac{L_1+L_2}{S} - {\mathbf{k}}_j^r \frac{L_2}{S}~,
\end{equation}
where $l$ is the pixel width, $h$ is the pixel height, $L_1$ is the distance between the grids, $L_2$ is the distance between the rear grid and the detector, $S$ is the distance between {\em{Solar Orbiter}} and the Sun (here we assume $S=1$ AU), while ${\mathbf{k}}_j^f$ and ${\mathbf{k}}_j^r$ are the wave vectors associated to the front and rear grid of detector $j$, respectively; $V({\boldsymbol{\xi}})$ is the visibility sampled at $(u,v)$ point ${\boldsymbol{\xi}}$, i.e.
\begin{equation}\label{visibility}
V({\boldsymbol{\xi}}) = \int_{\R^2} \phi({\mathbf{x}}) \exp(2 \pi i {\boldsymbol{\xi}} \cdot {\bf{x}}) d{\mathbf{x}},
\end{equation}
where $\phi({\mathbf{x}})$ is the incoming flux. Using this definition and equations (\ref{count1-a})-(\ref{csi}) leads to
\begin{equation}\label{visibility-approximated}
V({\boldsymbol{\xi}}_j) \simeq \frac{1}{4M_1} [(C_j - A_j) + i (D_j - B_j)] \exp\left(i\frac{\pi}{4}\right)~,
\end{equation}
which is the model for the visibility formation process in {\em{STIX}}. On the other hand, using in (\ref{count1-a})-(\ref{visibility}) the fact that
\begin{equation}\label{visibility-complex}
{\overline{V({\boldsymbol{\xi}})}} = V(-{\boldsymbol{\xi}})~,
\end{equation}
and some easy computation concerning complex numbers, leads to
\begin{equation}\label{count2-a}
A_j \simeq \int_{\R^2} \phi({\mathbf{x}}) \left(M_0 - 2 M_1 \cos\left(2\pi {\boldsymbol{\xi}}_j \cdot {\mathbf{x}} - \frac{\pi}{4}\right)\right) d{\mathbf{x}}~,
\end{equation}
\begin{equation}\label{count2-b}
B_j \simeq \int_{\R^2} \phi({\mathbf{x}}) \left(M_0 - 2 M_1 \sin\left(2\pi {\boldsymbol{\xi}}_j \cdot {\mathbf{x}} - \frac{\pi}{4}\right)\right) d{\mathbf{x}}~,
\end{equation}
\begin{equation}\label{count2-c}
C_j \simeq \int_{\R^2} \phi({\mathbf{x}}) \left(M_0 + 2 M_1 \cos\left(2\pi {\boldsymbol{\xi}}_j \cdot {\mathbf{x}} - \frac{\pi}{4}\right)\right) d{\mathbf{x}}~,
\end{equation}
and
\begin{equation}\label{count2-d}
D_j \simeq \int_{\R^2} \phi({\mathbf{x}}) \left(M_0 + 2 M_1 \sin\left(2\pi {\boldsymbol{\xi}}_j \cdot {\mathbf{x}} - \frac{\pi}{4}\right)\right) d{\mathbf{x}}~.
\end{equation}
These latter four equations can be discretized in the Sun space, where $\phi({\mathbf{x}})$ is defined, and such discretization leads to
\begin{equation}\label{count3-a}
A_j \simeq \sum_{h=0}^{N_1-1} \sum_{k=0}^{N_2-1} \phi_{hk} \left(M_0 - 2 M_1 \cos\left(2\pi {\boldsymbol{\xi}}_j \cdot {\mathbf{x}}_{hk} - \frac{\pi}{4}\right)\right) \Delta x_1 \Delta x_2~,
\end{equation}
\begin{equation}\label{count3-b}
B_j \simeq \sum_{h=0}^{N_1-1} \sum_{k=0}^{N_2-1} \phi_{hk} \left(M_0 - 2 M_1 \sin\left(2\pi {\boldsymbol{\xi}}_j \cdot {\mathbf{x}}_{hk} - \frac{\pi}{4}\right)\right) \Delta x_1 \Delta x_2~,
\end{equation}
\begin{equation}\label{count3-c}
C_j \simeq \sum_{h=0}^{N_1-1} \sum_{k=0}^{N_2-1} \phi_{hk} \left(M_0 + 2 M_1 \cos\left(2\pi {\boldsymbol{\xi}}_j \cdot {\mathbf{x}}_{hk} - \frac{\pi}{4}\right)\right) \Delta x_1 \Delta x_2~,
\end{equation}
and
\begin{equation}\label{count3-d}
D_j \simeq \sum_{h=0}^{N_1-1} \sum_{k=0}^{N_2-1} \phi_{hk} \left(M_0 + 2 M_1 \sin\left(2\pi {\boldsymbol{\xi}}_j \cdot {\mathbf{x}}_{hk} - \frac{\pi}{4}\right)\right) \Delta x_1 \Delta x_2~,
\end{equation}
where
\begin{equation}\label{phi-discretized}
\phi_{hk} = \phi(x_h,x_k)~,
\end{equation}
\begin{equation}\label{discretization-1}
x_h = -X_1 + h \Delta x_1~~~h=0,\ldots,N_1-1~,
\end{equation}
\begin{equation}\label{discretization-2}
x_k = -X_2 + k \Delta x_2~~~k=0,\ldots,N_2-1~,
\end{equation}
\begin{equation}\label{discretization-3}
{\mathbf{x}}_{hk} = (x_h,x_k)~,
\end{equation}
$\Delta x_1 = 2X_1/N_1$, $\Delta x_2 = 2 X_2/N_2$, and $[-X_1,X_1] \times [-X_2,X_2]$ is the field-of-view (FOV) (here we have assumed that the map center is $(0,0)$). Equations (\ref{count3-a})-(\ref{discretization-3}) prove that the link between the photon flux incident into the telescope and the number of counts recorded by its pixels can be modeled by means of the standard linear imaging equation
\begin{equation}\label{imaging-equation}
H\Phi = c,
\end{equation}
where $\Phi\in\mathbb{R}^{N_1 N_2}$ is the discrete photon flux vector obtained by means of a lexicographical re-ordering of the $N_1\times N_2$ image to restore, $c\in\mathbb{R}^{120}$ is the vector of counts recorded in the $4$ pixels of all $30$ detectors and $H\in\mathbb{R}^{120\times N_1 N_2}$ is the matrix whose rows represent the transmission functions through the grid pair down to the detector pixels. In order to explicitly write the entries of this matrix in a more compact form, we observe that, in equation (\ref{count3-a})
\begin{equation}\label{cosine-1}
- \cos\left(2\pi {\boldsymbol{\xi}}_j \cdot {\mathbf{x}}_{hk} - \frac{\pi}{4}\right) = \cos\left(2\pi {\boldsymbol{\xi}}_j \cdot {\mathbf{x}}_{hk} + \frac{3\pi}{4}\right)~;
\end{equation}
in equation (\ref{count3-b})
\begin{equation}\label{cosine-2}
- \sin\left(2\pi {\boldsymbol{\xi}}_j \cdot {\mathbf{x}}_{hk} - \frac{\pi}{4}\right) = \cos\left(2\pi {\boldsymbol{\xi}}_j \cdot {\mathbf{x}}_{hk} + \frac{\pi}{4}\right)~;
\end{equation}
and, in equation (\ref{count3-d})
\begin{equation}\label{cosine-3}
\sin\left(2\pi {\boldsymbol{\xi}}_j \cdot {\mathbf{x}}_{hk} - \frac{\pi}{4}\right) = \cos\left(2\pi {\boldsymbol{\xi}}_j \cdot {\mathbf{x}}_{hk} - \frac{3\pi}{4}\right)~.
\end{equation}
Therefore the entries of $H$ can be written as
\begin{equation}\label{imaging-matrix}
	H_{q,p} = \left[M_0 + 2 M_1\cos\left(2\pi \boldsymbol{\xi}_j \cdot \mathbf{x}_{p} + \frac{5 \pi}{4} -l\frac{\pi}{2} \right)\right]\Delta x_1 \Delta x_2  ~,
\end{equation}
where $q=(j, l)$, $j=1, \dots, 30$ is the detector number, $l=1, \dots, 4$ is the pixel number, $p=1, \dots, N_1  N_2$ and $\mathbf{x}_{p}$ is the result of the re-ordering of $\mathbf{x}_{hk}$. Equations (\ref{imaging-equation}), (\ref{imaging-matrix}) represent the model for the count formation process in {\em{STIX}}. The availability of such model allows the use and formulation of image reconstruction methods taking as input the count vectors recorded by {\em{STIX}} pixels.

\section{Expectation Maximization}
\begin{figure}[h]
\includegraphics[width=.33\textwidth]{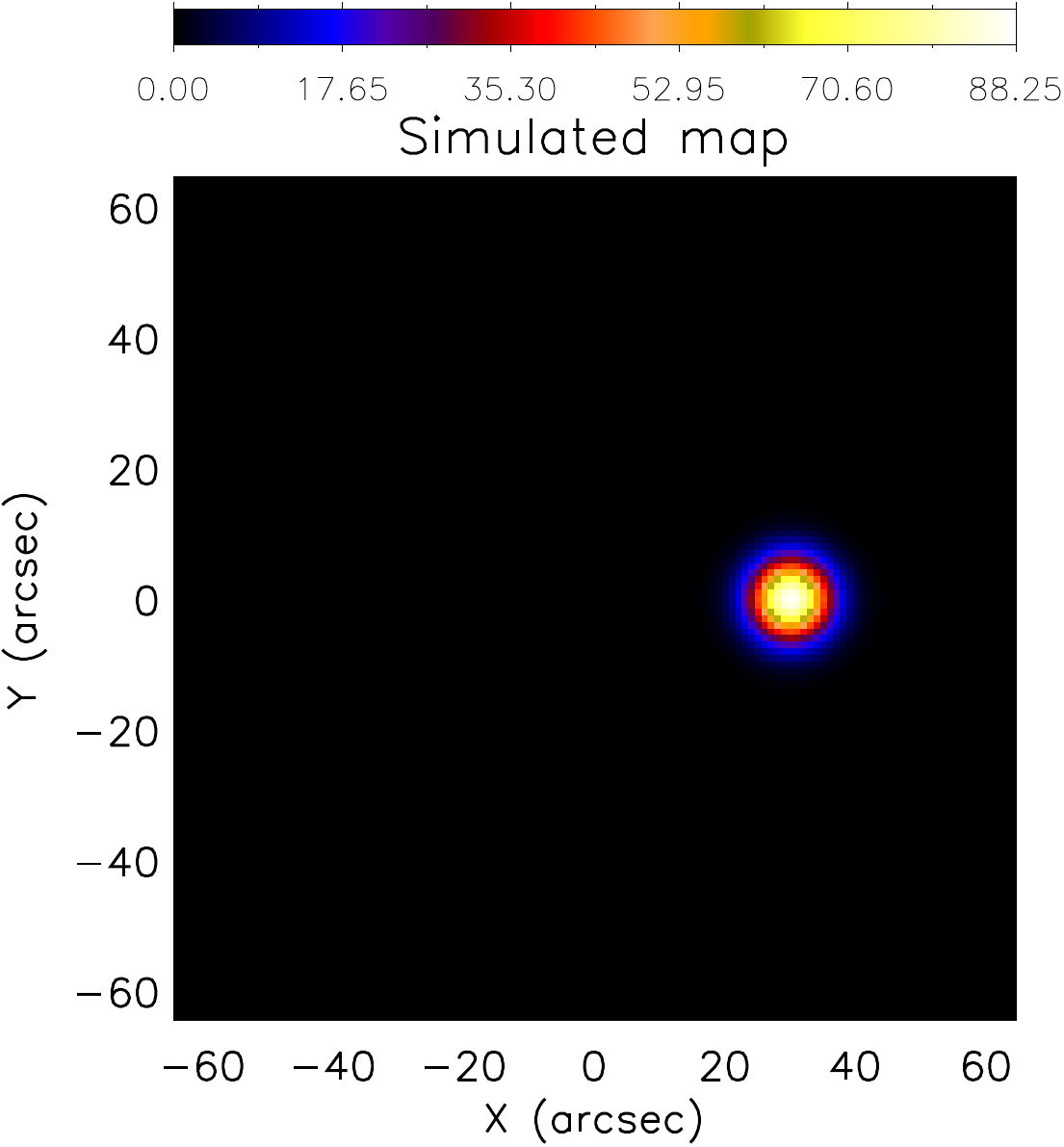} \hskip1cm
\includegraphics[width=.5\textwidth]{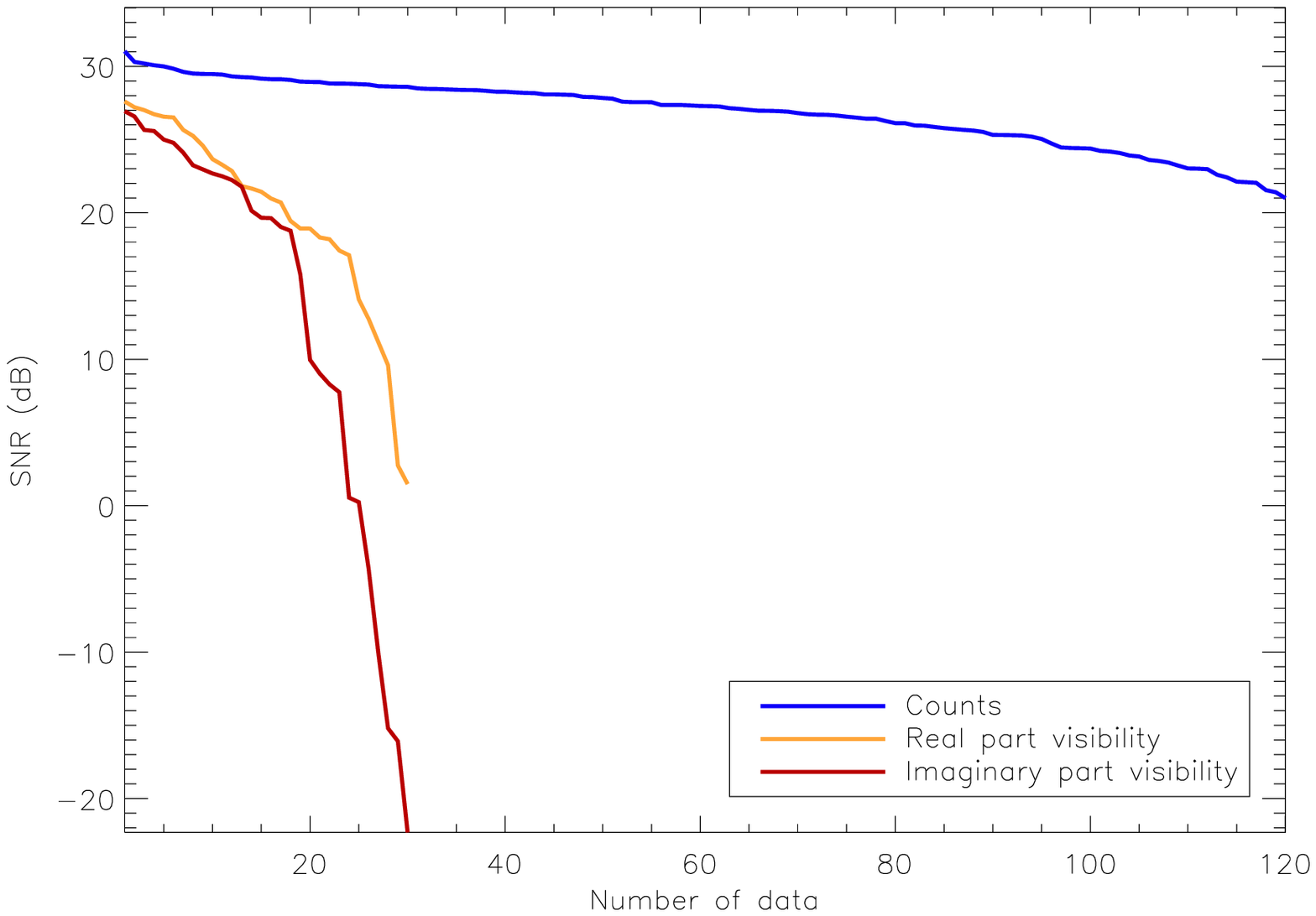}
\caption{Count statistic versus visibility statistic in the case of $50$ random realizations of the count vector produced by the {\em{STIX}} DPS. Left panel: simulated source configuration; right panel: signal-to-noise-ration (SNR) associated to each data component. Specifically, the blue line represents the SNR of each count vector component; the orange line represents the SNR of the real part of each visibility; the red line represents the SNR of the imaginary part of each visibility. The SNR values are ordered from the highest to the smallest one.}\label{fig:figure-1}
\end{figure}

The use of the count-based framework (\ref{imaging-equation}), (\ref{imaging-matrix}) instead of the visibility-based Fourier model (\ref{visibility}), (\ref{visibility-approximated}) has two immediate advantages. The first one is that in (\ref{imaging-equation}), (\ref{imaging-matrix}) data are more numerous, since counts are not mixed up (i.e., subtracted) in order to obtain visibilities, as in (\ref{visibility}), (\ref{visibility-approximated}). The second advantage is that the signal-to-noise ratio in problem (\ref{imaging-equation}), (\ref{imaging-matrix}) is higher than in (\ref{visibility}), (\ref{visibility-approximated}). Indeed, counts are Poisson variables, while visibilities are obtained by subtracting counts and therefore follow the Skellam statistics, in which the standard deviation is greater than the one associated to Poisson statistics. This is shown also empirically in Figure \ref{fig:figure-1}: we have constructed the vector $\Phi$ corresponding to the Gaussian source in the left panel and then we used the {\em{STIX}} DPS to randomly generate $50$ vectors $c$; using the components in $c$ we computed the corresponding $50$ realizations of the $30$ {\em{STIX}} visibilities. In the right panel of the figure, the signal-to-noise ratio of each count component is compared to the signal-to-noise ratio of the real and imaginary part of each visibility. 

Image reconstruction from {\em{STIX}} counts requires the numerical solution of equation (\ref{imaging-equation}), which suffers from numerical instability owing to the conditioning of matrix $H$. However, a stable and reliable reconstruction can be obtained by early stopping the Expectation Maximization (EM) iterative algorithm that exploits the Poisson nature of the noise affecting the components of vector $c$. Indeed, EM assumes that $c$ is the realization of a random variable $C$ with Poisson distribution so that the probability of observing $c$ when the incoming flux is represented by $\Phi$ is
\begin{equation}\label{likelihood}
p(c\vert H\Phi) = \prod_{q=1}^{120}\frac{\exp\left(-\left( H\Phi\right)_q \right)\left(H\Phi \right)_q^{c_q} }{c_q!}~.
\end{equation}
The constrained maximum likelihood problem addresses the optimization problem
\begin{equation}\label{constrained-maximum-likelihood}
\hat{\Phi} = \max_{\Phi \geq 0}p(c \vert H\Phi)~,
\end{equation}
where the positivity constraint is referred to each component of $\Phi$. It can be proven that (\ref{constrained-maximum-likelihood}) can be expressed as a fixed point problem, which can be solved by means of the successive approximation scheme
\begin{equation}\label{EM}
\Phi^{(n+1)} = \frac{\Phi^{(n)}}{H^{\text{T}} {\mathbf{1}}} \left(H^{\text{T}}\frac{c}{H\Phi^{(n)}} \right)~~,~~n=1,\ldots
\end{equation}
where ${\mathbf{1}}$ is the vector with $1$ for each component. The stopping rule introduced in \citet{2014InvPr..30c5012B} provides an approximate solution of (\ref{imaging-equation}) in the sense of asymptotic regularization \citep{be2017}, able to realize a trade-off between numerical stability and data fitting.

	\section{Results}
	
		\begin{figure}[h]
		\begin{center}
		\begin{tabular}{ccc}
				
				\includegraphics[width=.3\textwidth]{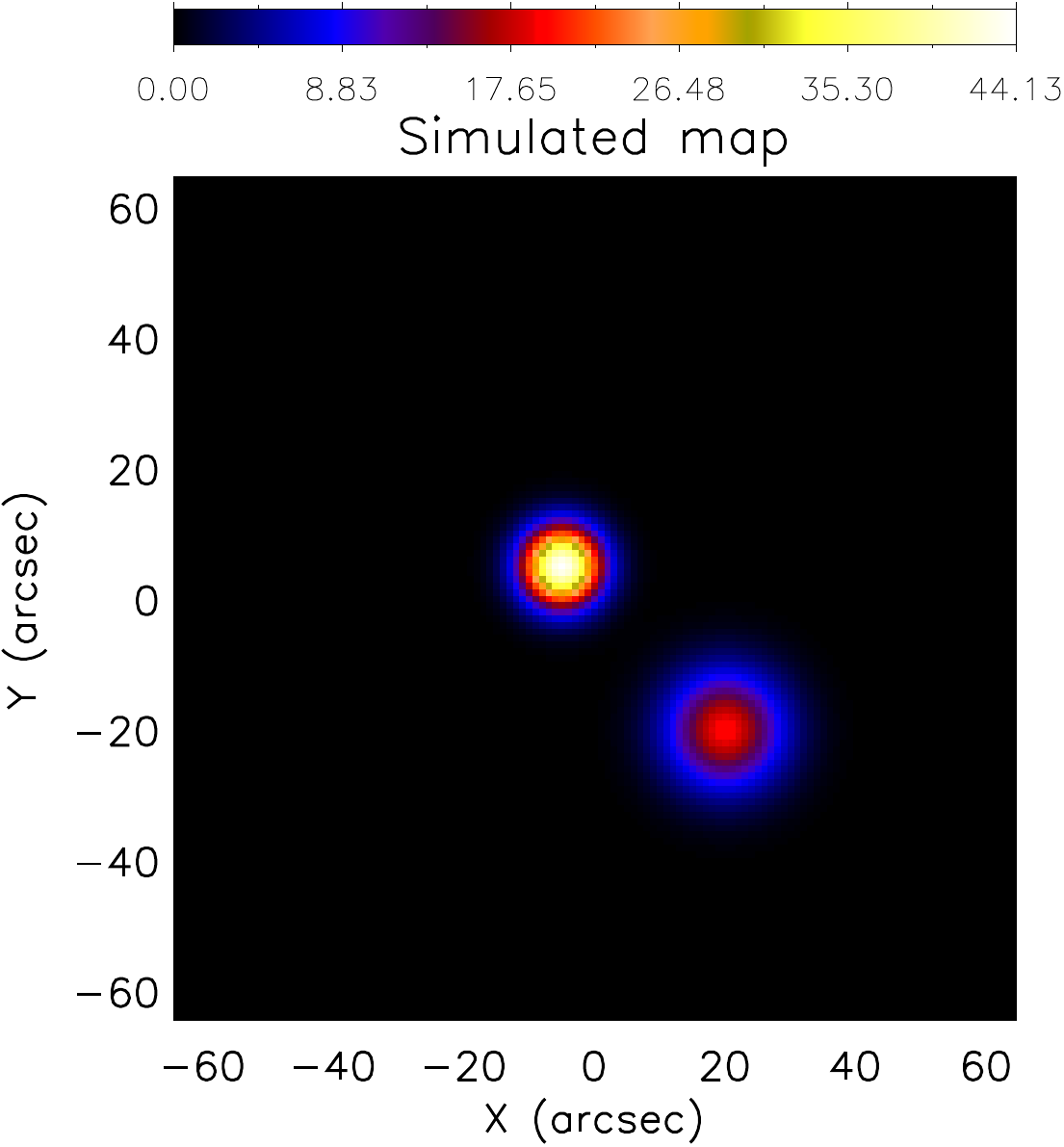} &
				\includegraphics[width=.3\textwidth]{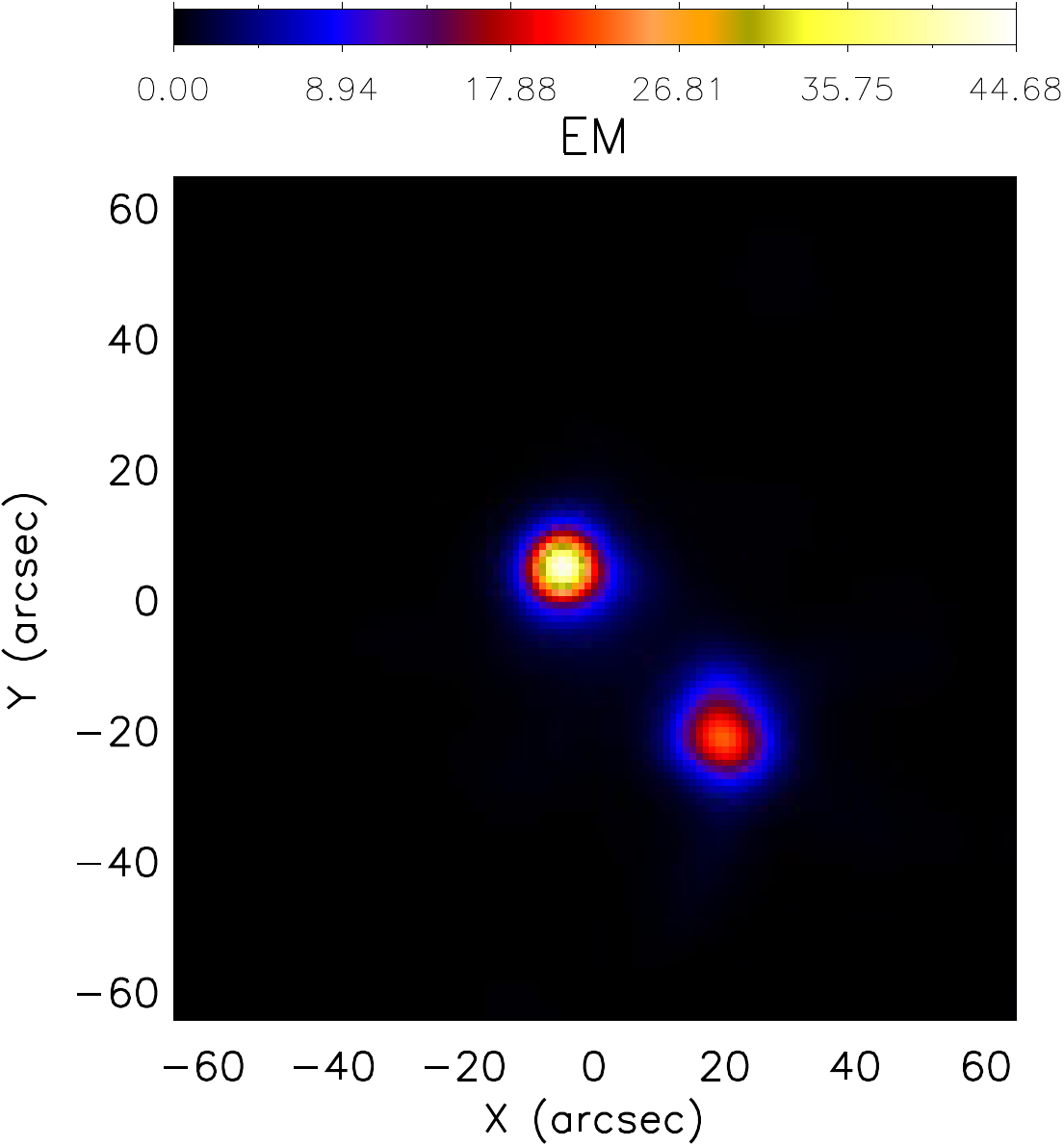} &
				\includegraphics[width=.3\textwidth]{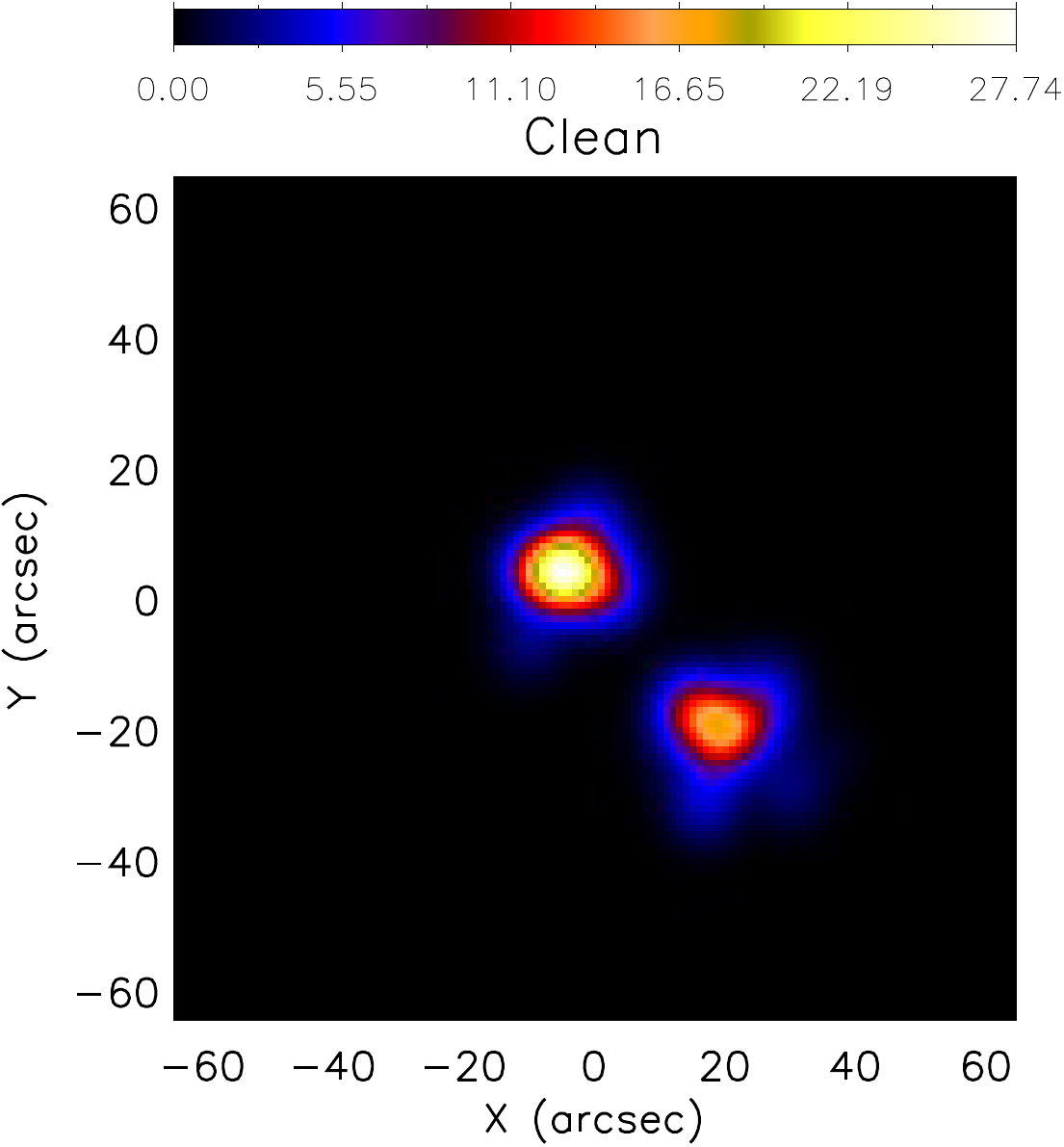} \\
			
				
				\includegraphics[width=.3\textwidth]{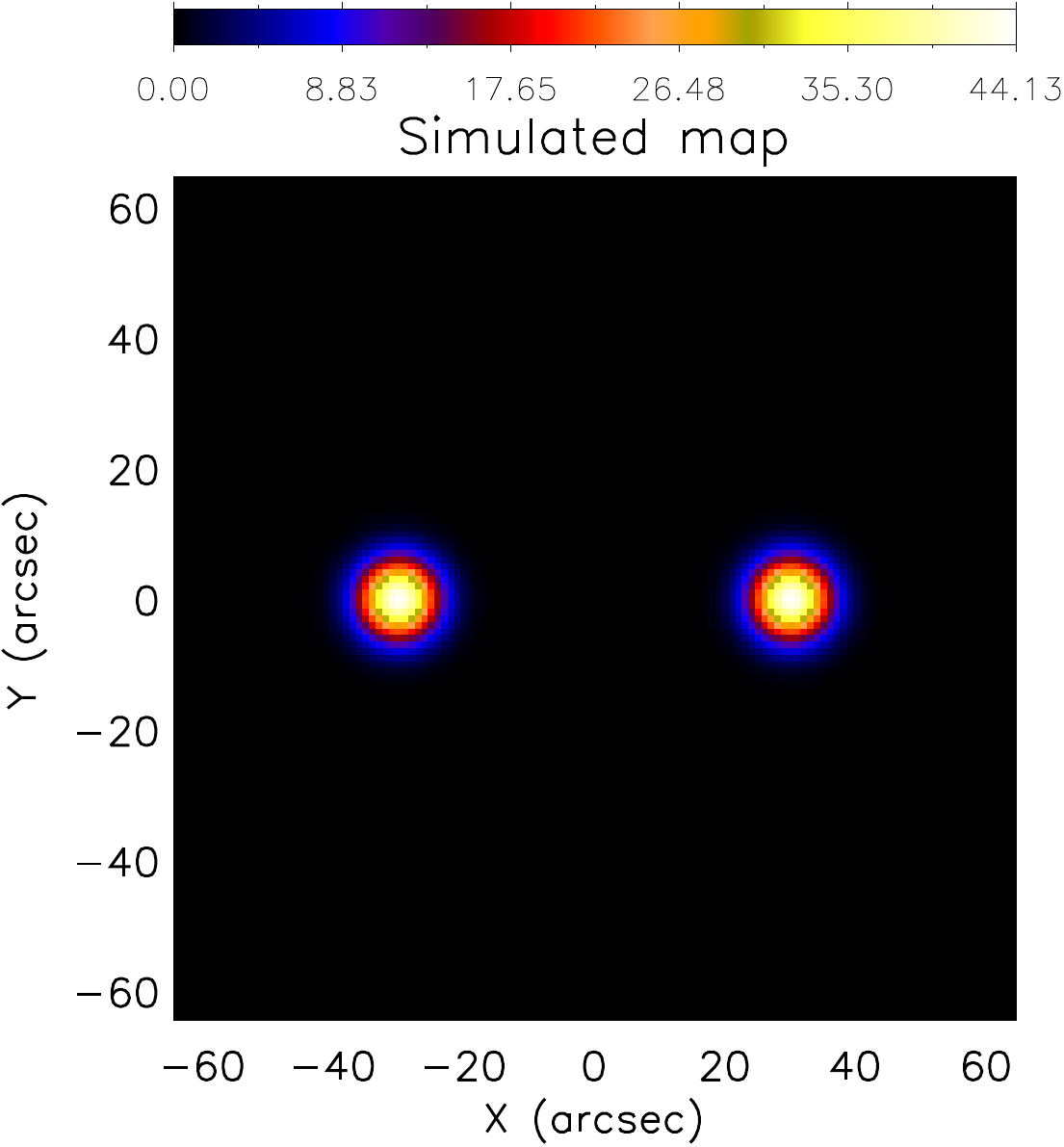} &
				\includegraphics[width=.3\textwidth]{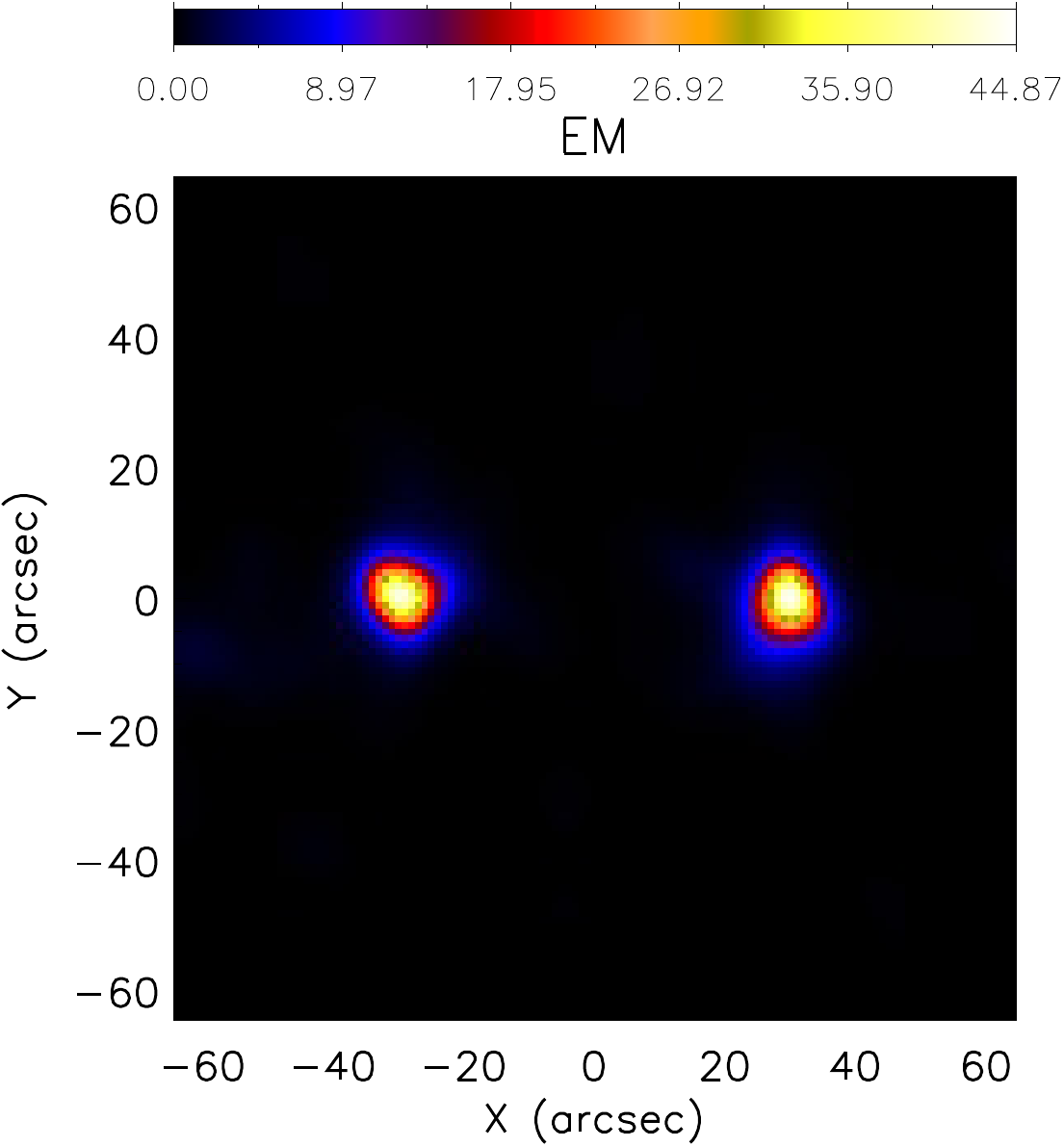} &
				\includegraphics[width=.3\textwidth]{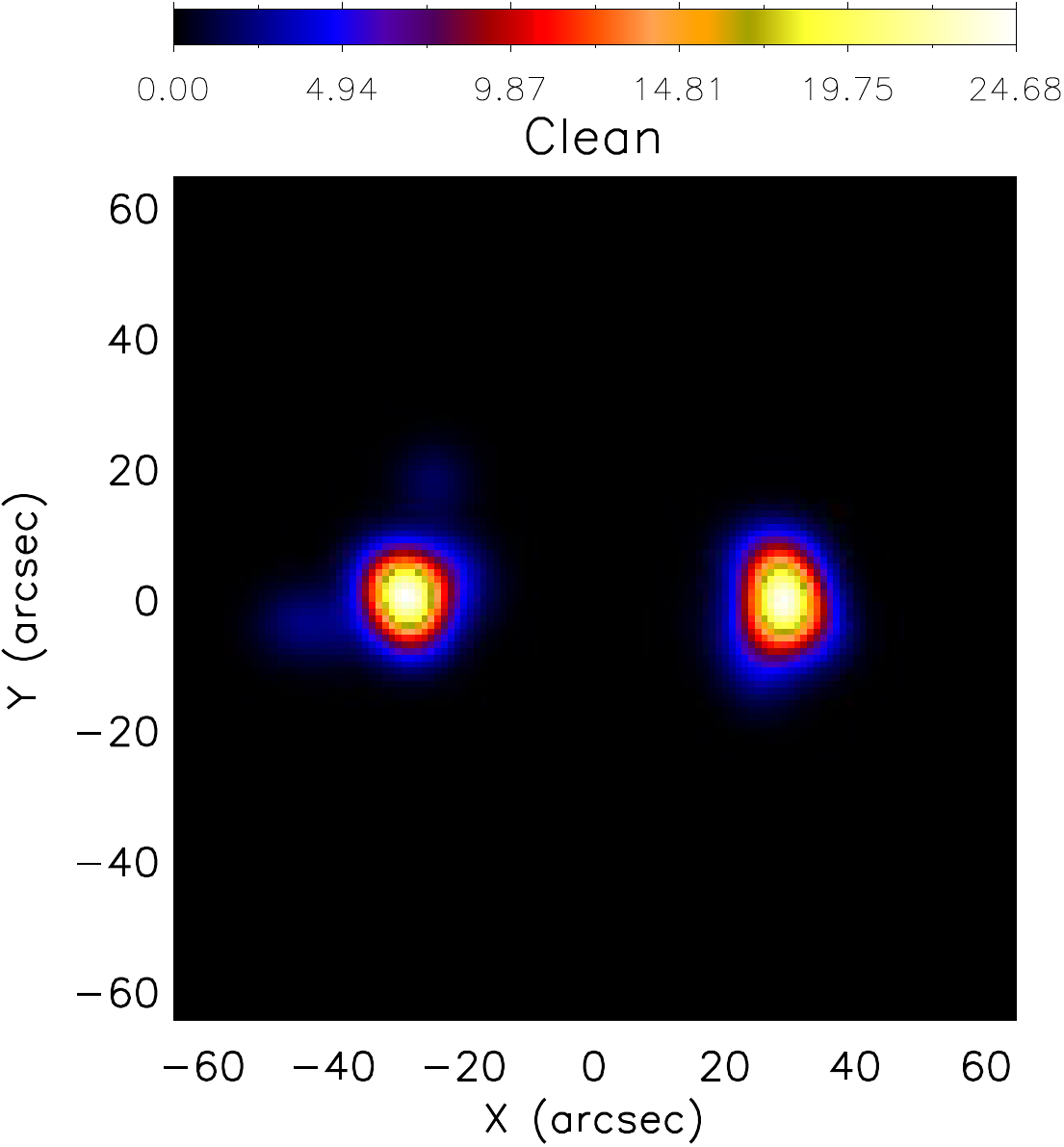} \\
			
				
				\includegraphics[width=.3\textwidth]{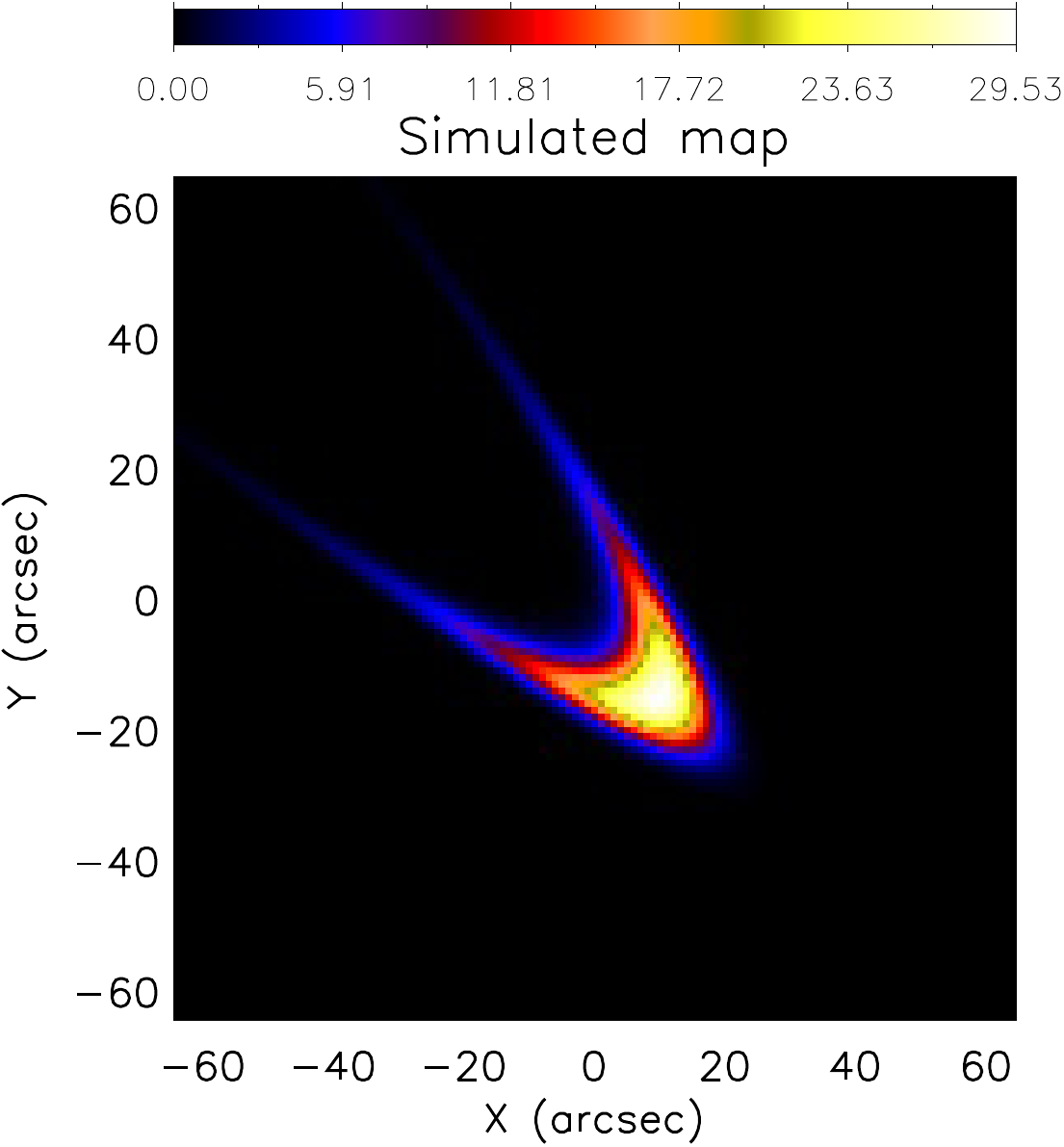} &
				\includegraphics[width=.3\textwidth]{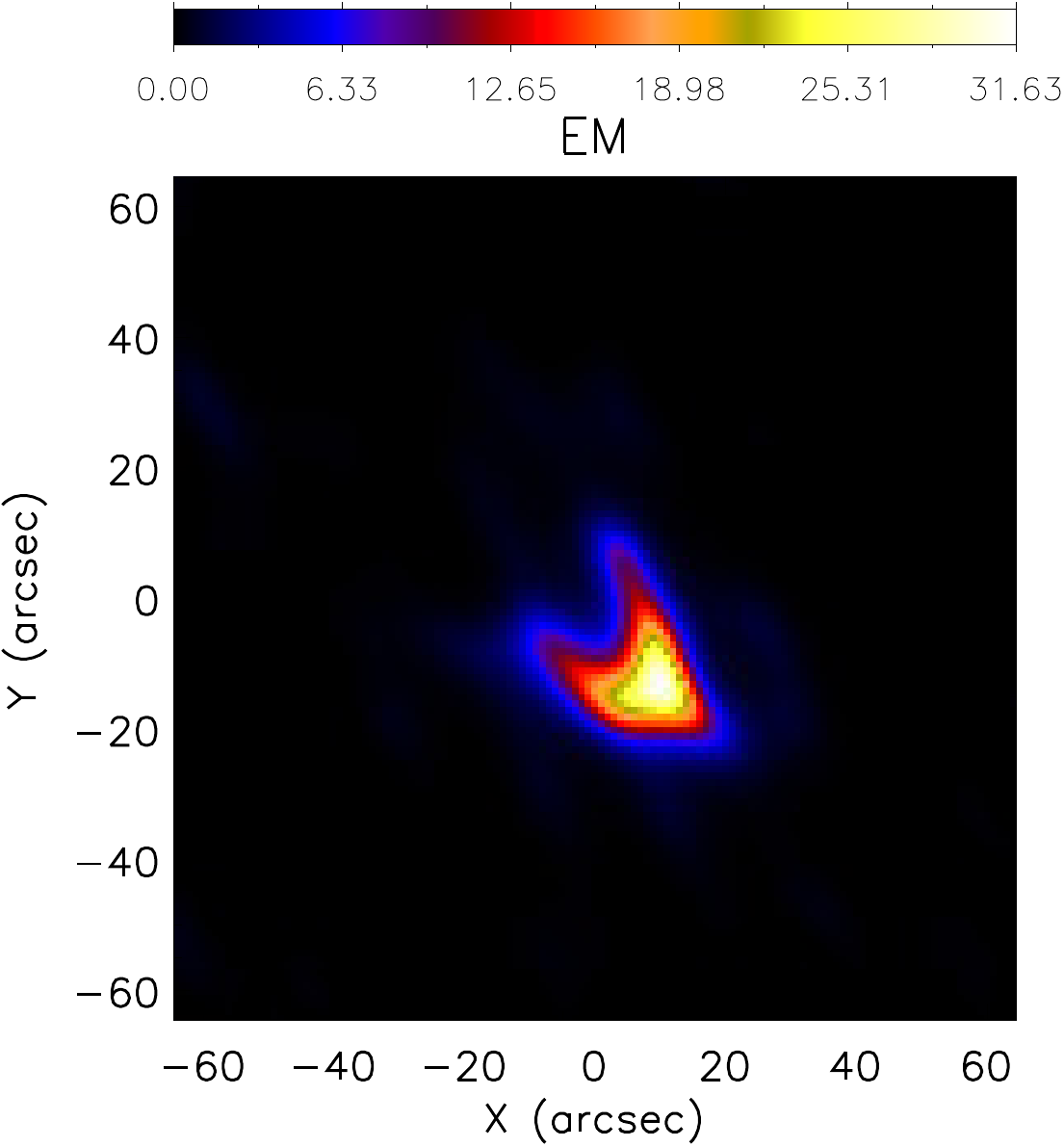} &
				\includegraphics[width=.3\textwidth]{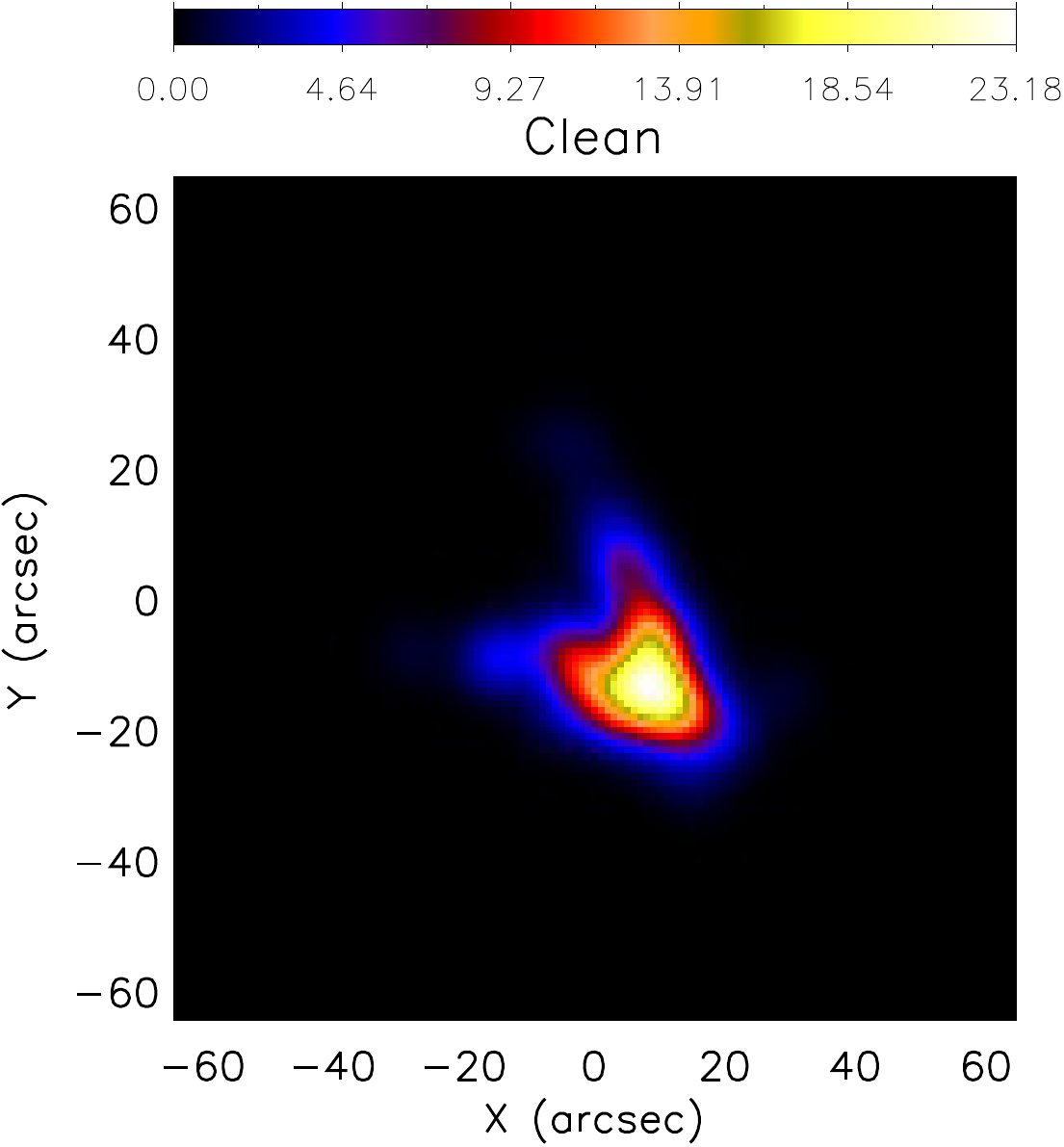} \\
			
				
				\includegraphics[width=.3\textwidth]{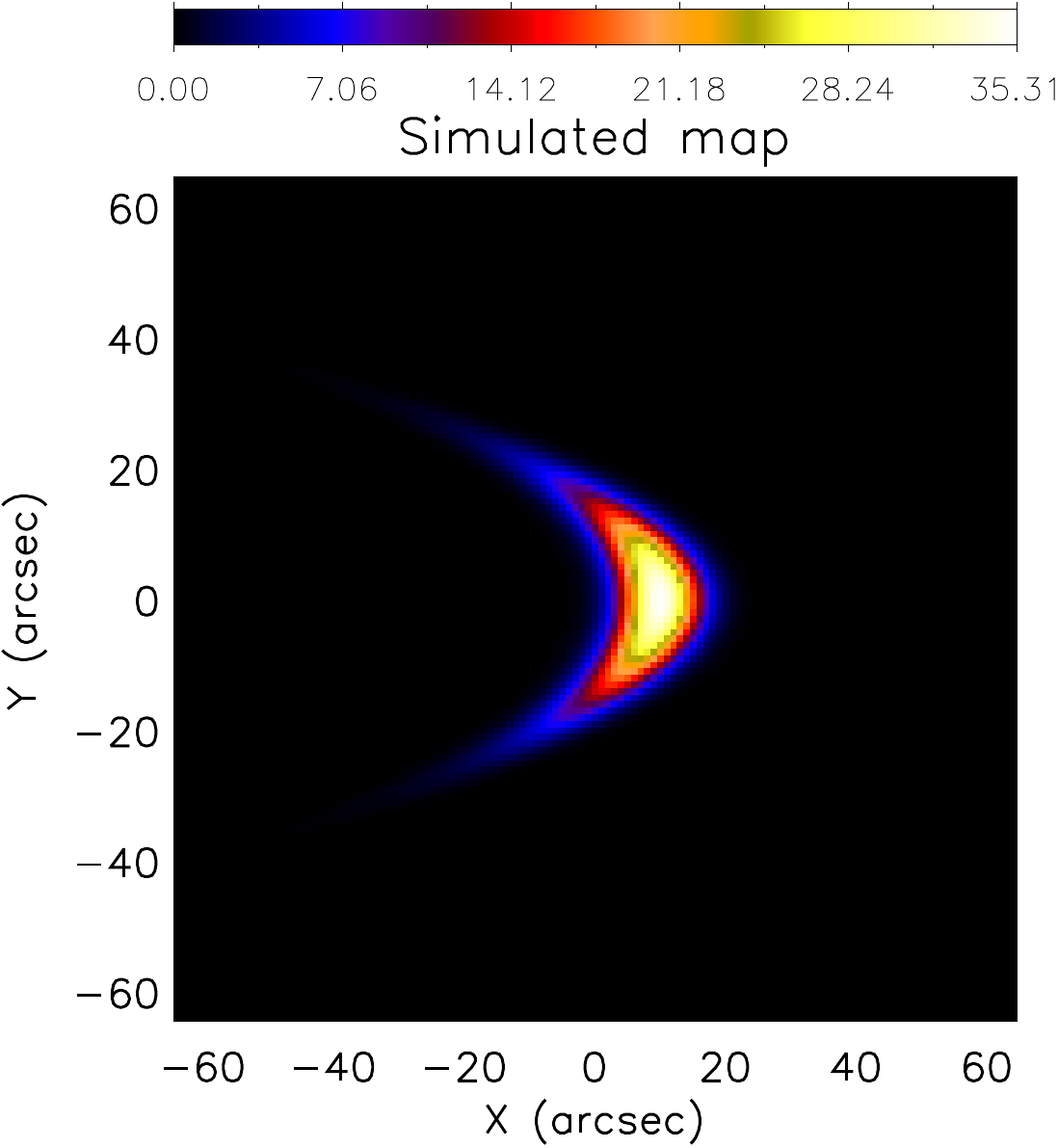} & 
				\includegraphics[width=.3\textwidth]{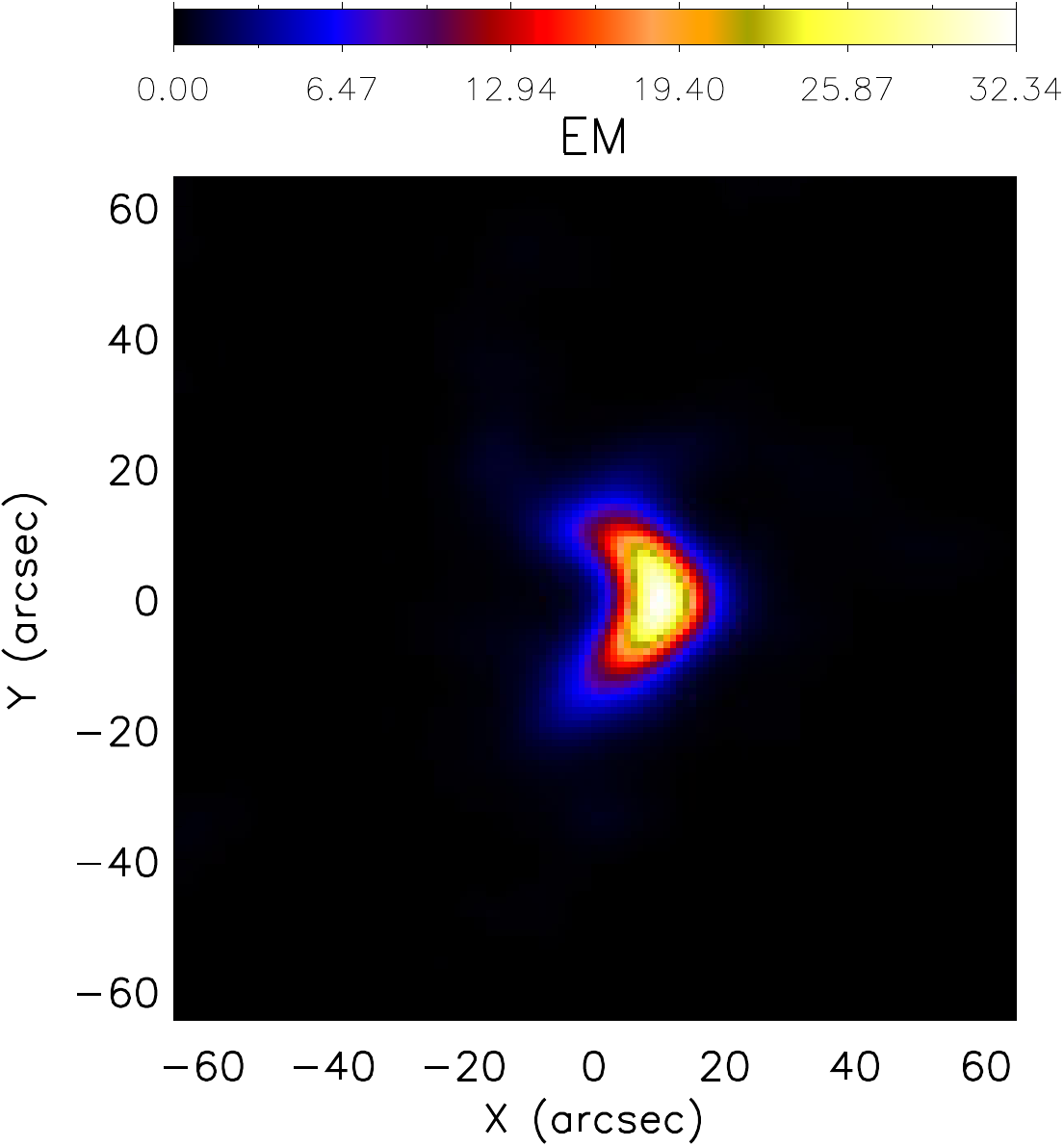} &
				\includegraphics[width=.3\textwidth]{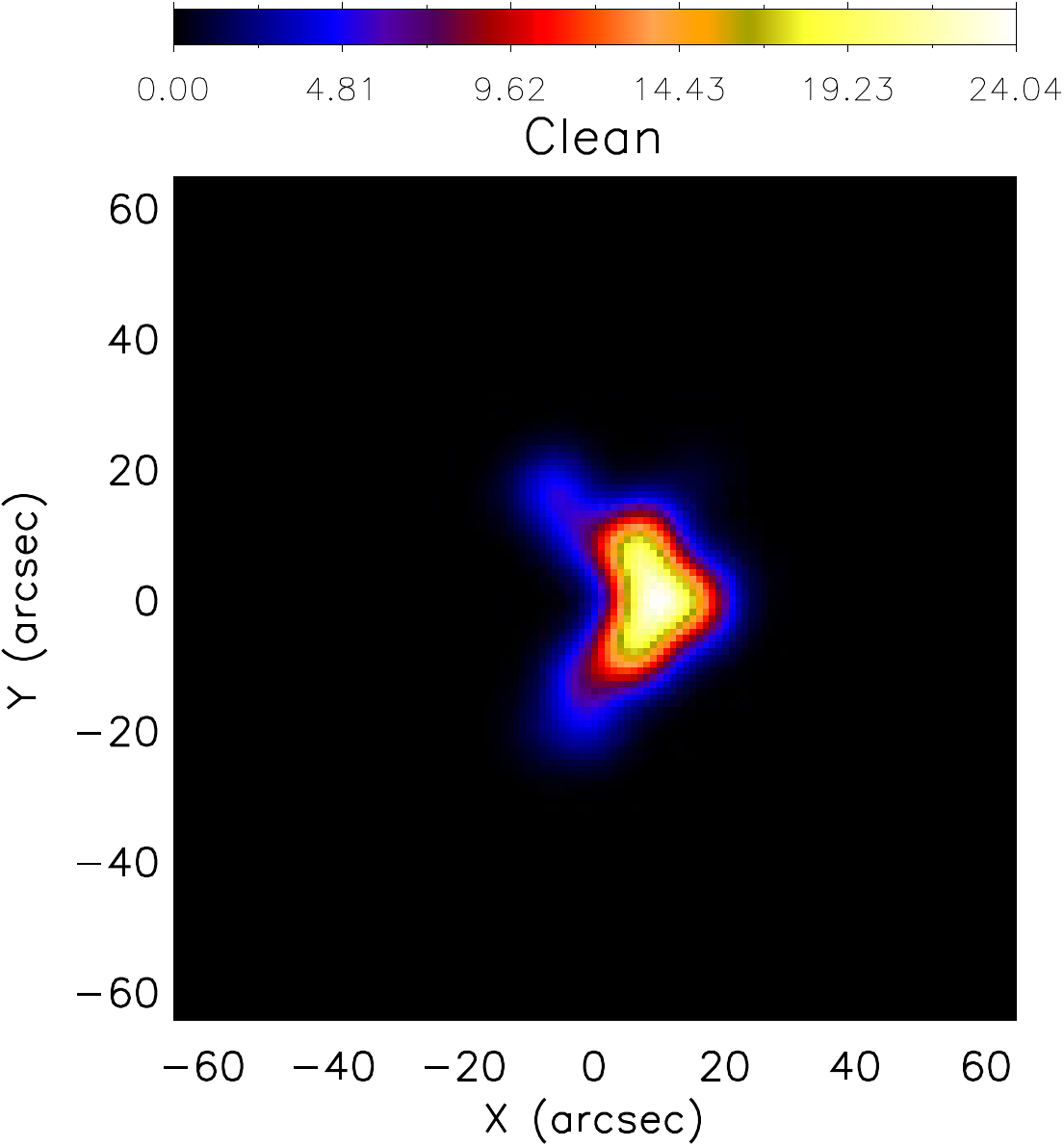}
						
		\end{tabular}
		\end{center}
			\caption{Reconstructions of four source configurations characterized by an overall incident flux of $10^4$ photon s$^{-1}$ cm$^{-2}$ (medium statistic). First column: simulated configurations (ground truth); second column: count-based EM reconstructions; third column: reconstructions provided by the visibility-based CLEAN algorithm.}\label{fig:figure-2}
		\end{figure}

\begin{table}[h]
	\centering	
	\resizebox{\linewidth}{!}{\begin{tabular}{rcccccccc} 
			\toprule
			\multicolumn{9}{c}{\textbf{configuration FF1}}\\
			\midrule
			&\multicolumn{3}{c}{First Peak}						&\multicolumn{3}{c}{Second Peak} 					 &Total flux ($\times 10^3$) &C-statistic\\
			\cmidrule(lr){2-4}
			\cmidrule(lr){5-7}
			
			&X			   &Y			 &FOHM ($\times 10^3$)	&X			   &Y		 		&FOHM ($\times 10^3$) &						&\\
			\cmidrule{2-9}
			Simulated	&-5.0		   &5.0			 &2.46					&20.0		   &-20.0	    	&2.50				  &$10.00$				&\\
			EM			&$-4.6\pm0.5$  &$5.0\pm 0.0$ &$1.86\pm 0.10$		&$20.1\pm1.4$ &$-21.1\pm 1.0$	&$1.89\pm 0.15$		  &$10.76\pm 0.04$		&$3.5\pm 0.2$\\
			Clean	&$-4.3\pm0.5$  &$4.3\pm 0.5$ &$1.84\pm 0.10$		&$19.6\pm 1.1$ &$-19.3\pm 0.7$ 	&$1.63\pm 0.16$		  &$8.38\pm0.11$		&$35.4\pm 4.0$\\
			\midrule
			\midrule
			\multicolumn{9}{c}{\textbf{configuration FF2}}\\
			\midrule
			&\multicolumn{3}{c}{First Peak}					&\multicolumn{3}{c}{Second Peak} 					&Total flux($\times 10^3$) &C-statistic\\
			\cmidrule(lr){2-4}
			\cmidrule(lr){5-7}
			
			&X				&Y				&FOHM ($\times 10^3$) &X			 &Y				&FOHM ($\times 10^3$) &						&\\
			\cmidrule{2-9}
			Simulated	&-30.0			&0.0			&2.55				  &30.0			 &0.0			&2.55				  &$10.00$				&\\
			EM			&$-29.9\pm 0.7$	&$0.4\pm 0.7$	&$1.78\pm 0.10$		  &$29.8\pm 0.4$ &$-0.6\pm 0.5$	&$1.79\pm 0.11$		  &$10.66\pm 0.03$		&$3.1\pm 0.1$\\
			Clean	&$-29.0\pm 0.0$	&$0.5\pm 0.5$	&$1.82\pm 0.07$		  &$28.9\pm 0.3$ &$-0.6\pm 0.5$	&$1.78\pm 0.06$		  &$8.08\pm 0.10$		&$41.2\pm 2.8$\\
			\midrule
			\midrule
			\multicolumn{9}{c}{\textbf{configuration LF1}}\\
			\midrule
			&\multicolumn{3}{c}{Peak}									&Total flux ($\times 10^3$)			&C-statistic  	&	&	&\\
			\cmidrule(lr){2-4}
			&X				&Y					&FOHM ($\times 10^3$) 	&									&			  	&	&	&\\
			\cmidrule{2-9}
			Simulated	&10.0			&-15.0				&4.96					&10.00								&			 	&	&	&\\
			EM			&$9.9\pm 0.7$	&$-14.0 \pm 1.2$	&$3.53\pm 0.50$			&$10.67 \pm 0.04$					&$3.3 \pm 0.3$	& 	&	&\\
			Clean	&$8.5\pm 0.7$	&$-14.1 \pm 0.7$	&$3.51\pm 0.23$			&$8.39 \pm 0.08$					&$32.9 \pm 2.4$	& 	&	&\\
			\midrule
			\midrule
			\multicolumn{9}{c}{\textbf{configuration LF2}}\\
			\midrule
			&\multicolumn{3}{c}{Peak}									&Total flux ($\times 10^3$)				&C-statistic  	&	&	&\\
			\cmidrule(lr){2-4}
			&X				&Y					&FOHM ($\times 10^3$)	&										&			  	&	&	&\\
			\cmidrule{2-9}
			Simulated	&10.0			&0.0				&5.04					&10.00									&			 	&	&	&\\
			EM			&$9.3\pm 0.7$	&$-1.6 \pm 1.8$		&$3.59 \pm 0.62$		&$10.73 \pm 0.04$							&$3.4 \pm 0.2$	& 	&	&\\
			Clean	&$8.9\pm 0.3$	&$-1.6 \pm 1.4$		&$3.96 \pm 0.23$		&$8.63 \pm 0.12$						&$27.7 \pm 2.5$	& 	&	&\\
			\bottomrule	
			\end{tabular}}
			\caption{Reconstruction of four source configurations characterized by an overall incident photon flux of $10^4$ photons cm$^{-2}$ s $^{-1}$ (medium statistic). The  morphological and photometric parameters reconstructed by EM are compared with the ground truth and with the values provided by CLEAN (with FOHM we denote the integrated Flux Over Half Maximum, i.e. the flux above $50\%$ level). The positions are in arcsec; the total flux and the FOHM are in photons cm$^{-2}$ s$^{-1}$.}\label{table:table-1}
		\end{table}

	\begin{figure}[h]
		\begin{center}
		\begin{tabular}{ccc}
			
				\includegraphics[width=.3\textwidth]{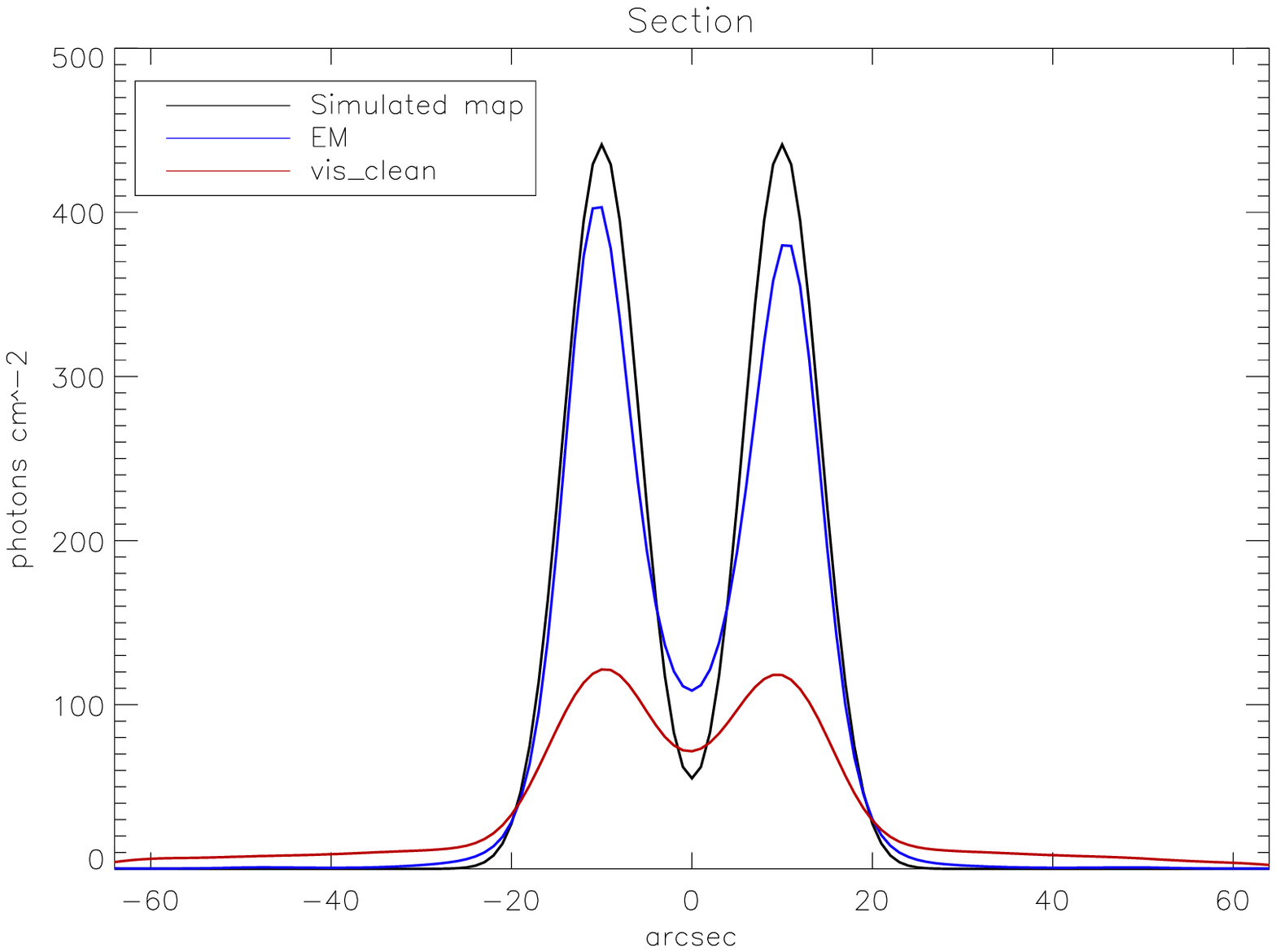} &
				\includegraphics[width=.3\textwidth]{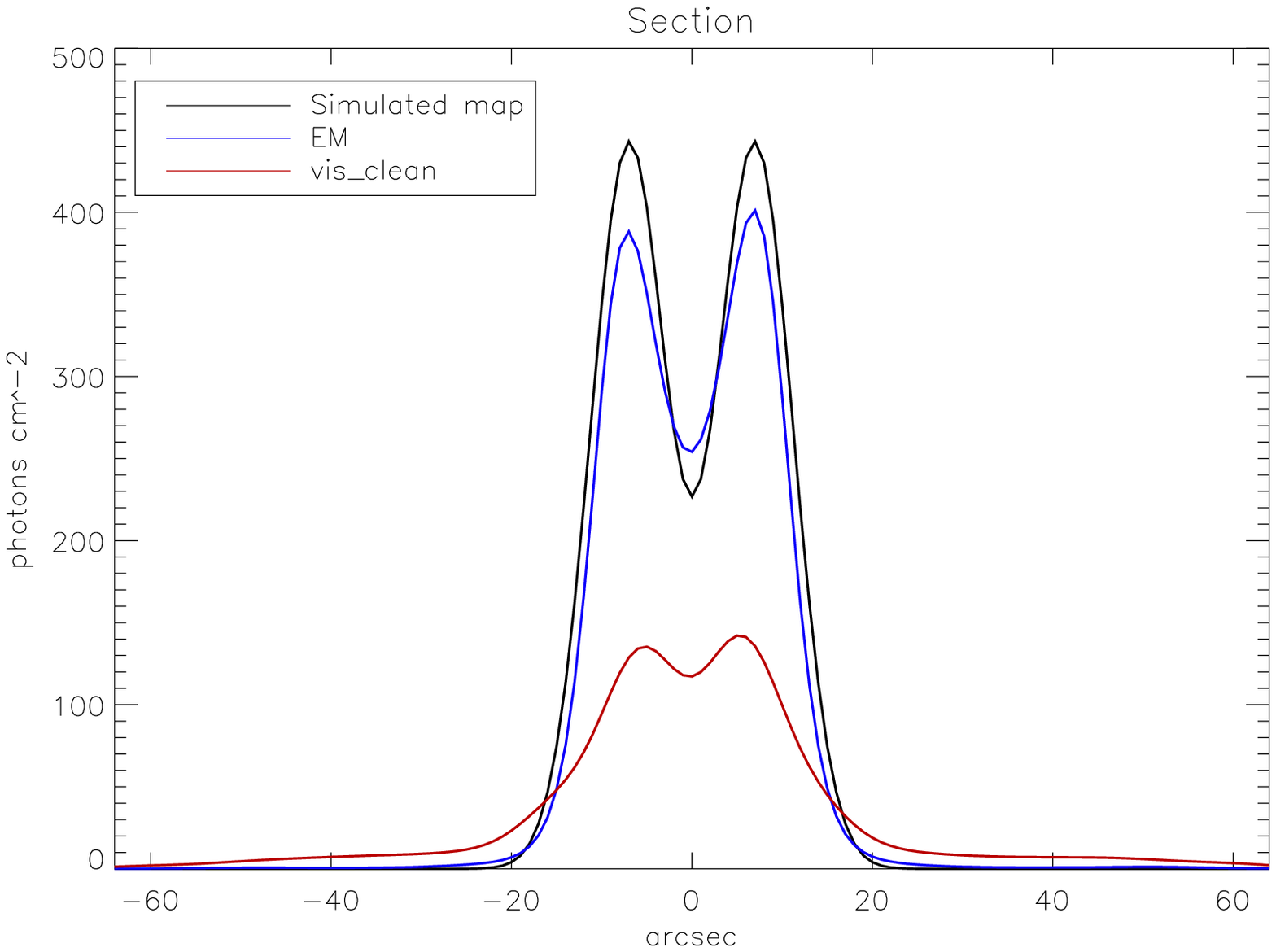} &
				\includegraphics[width=.3\textwidth]{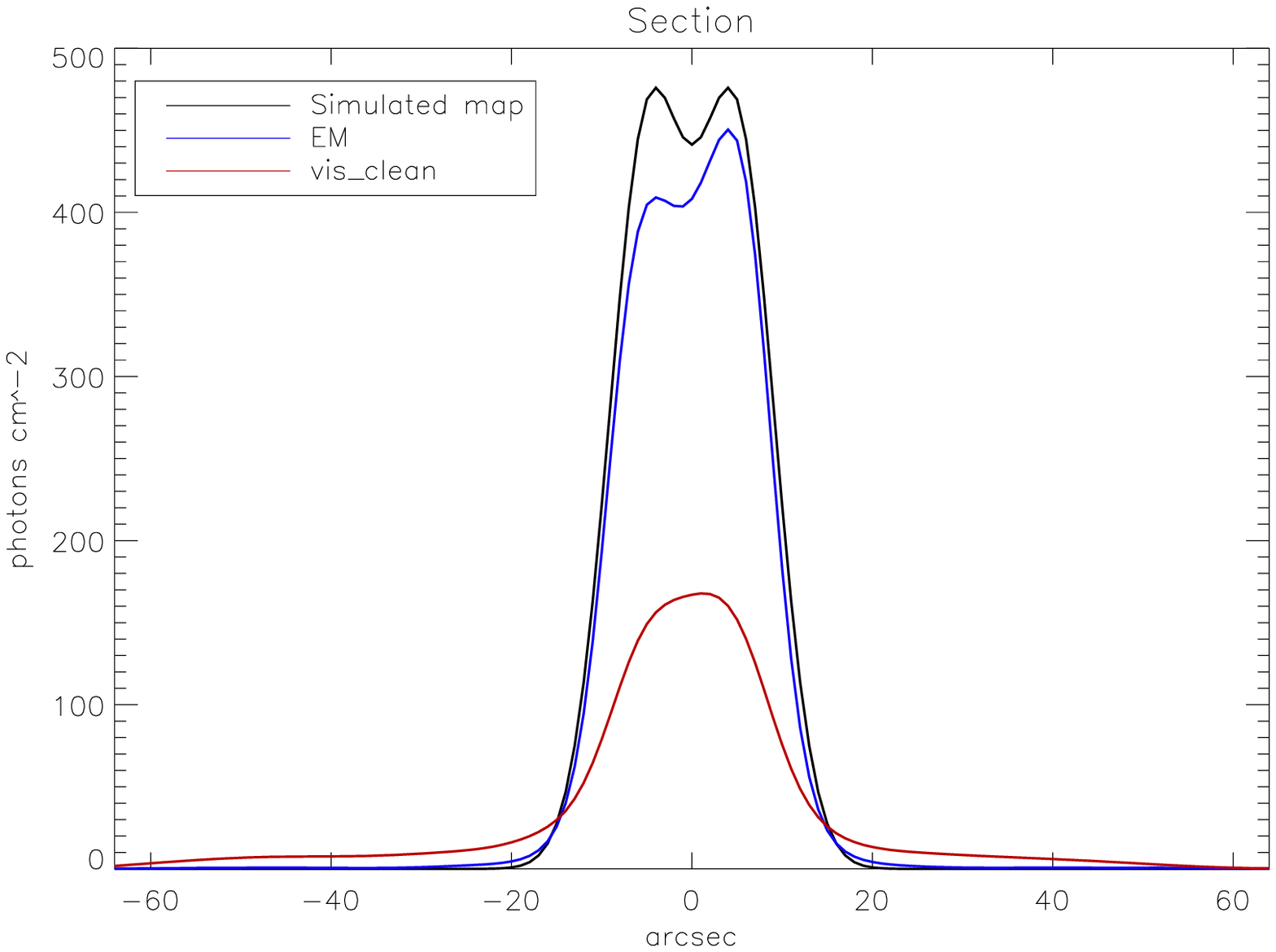}	 \\
			
			
				\includegraphics[width=.3\textwidth]{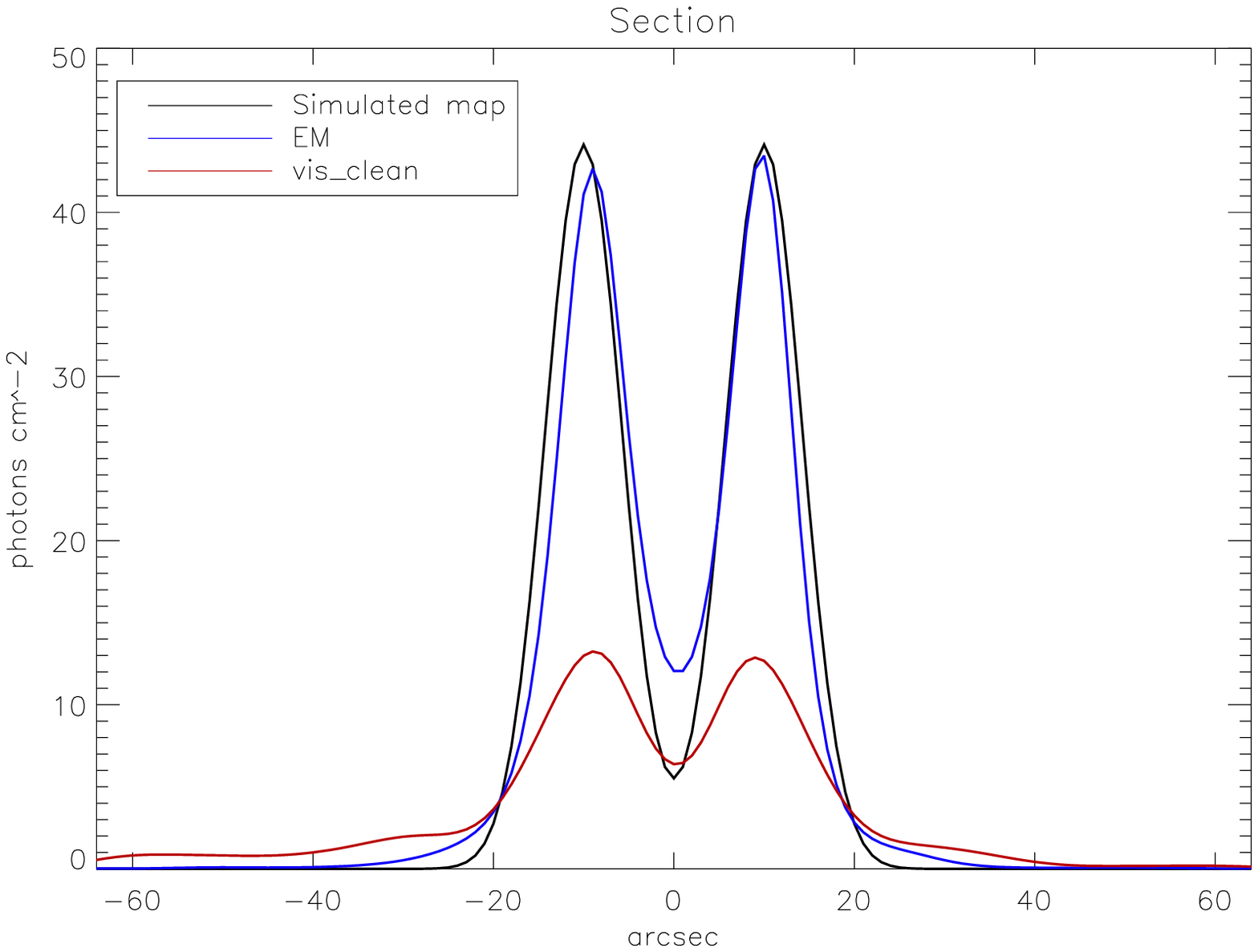} &
				\includegraphics[width=.3\textwidth]{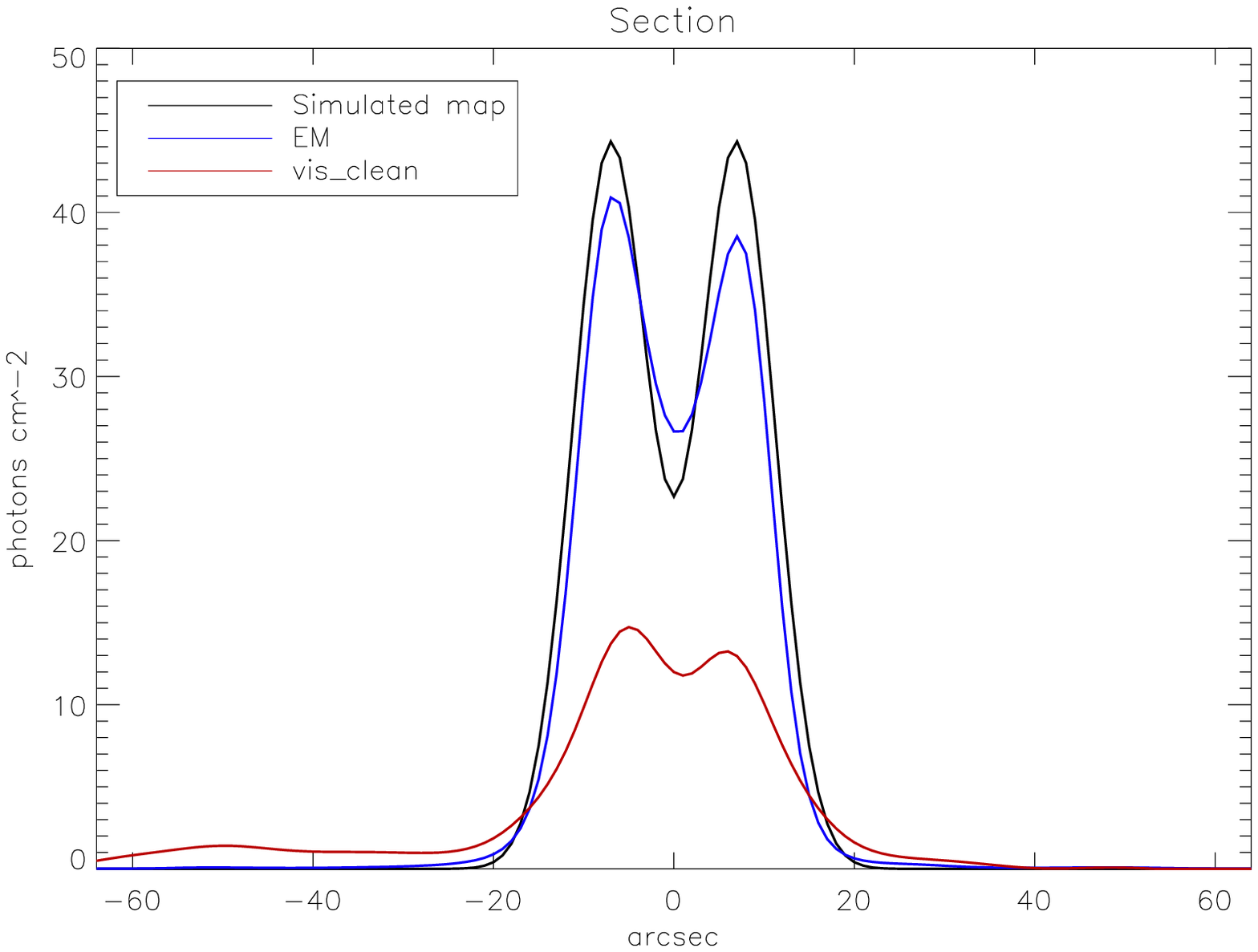} &
				\includegraphics[width=.3\textwidth]{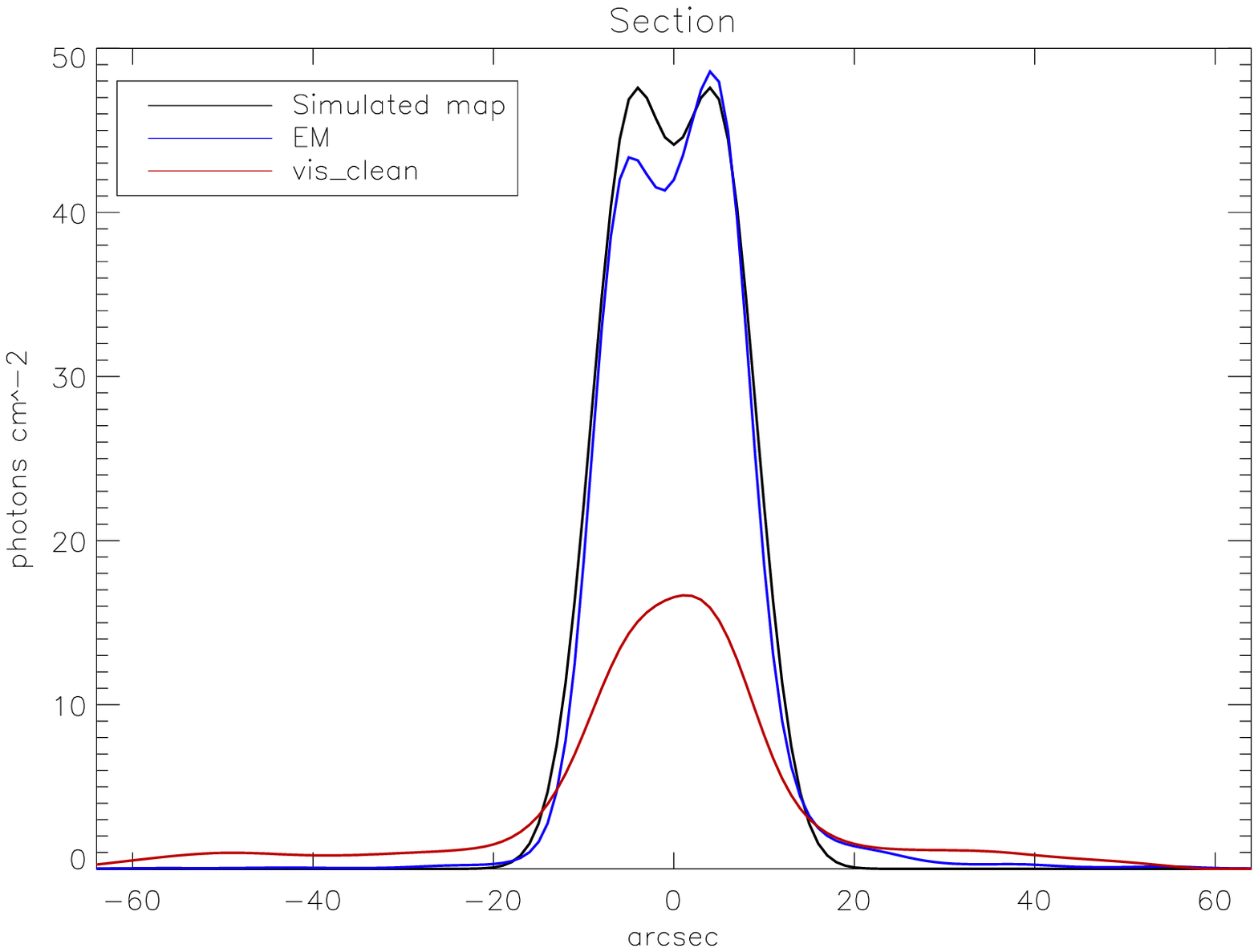} \\
			
			
				\includegraphics[width=.3\textwidth]{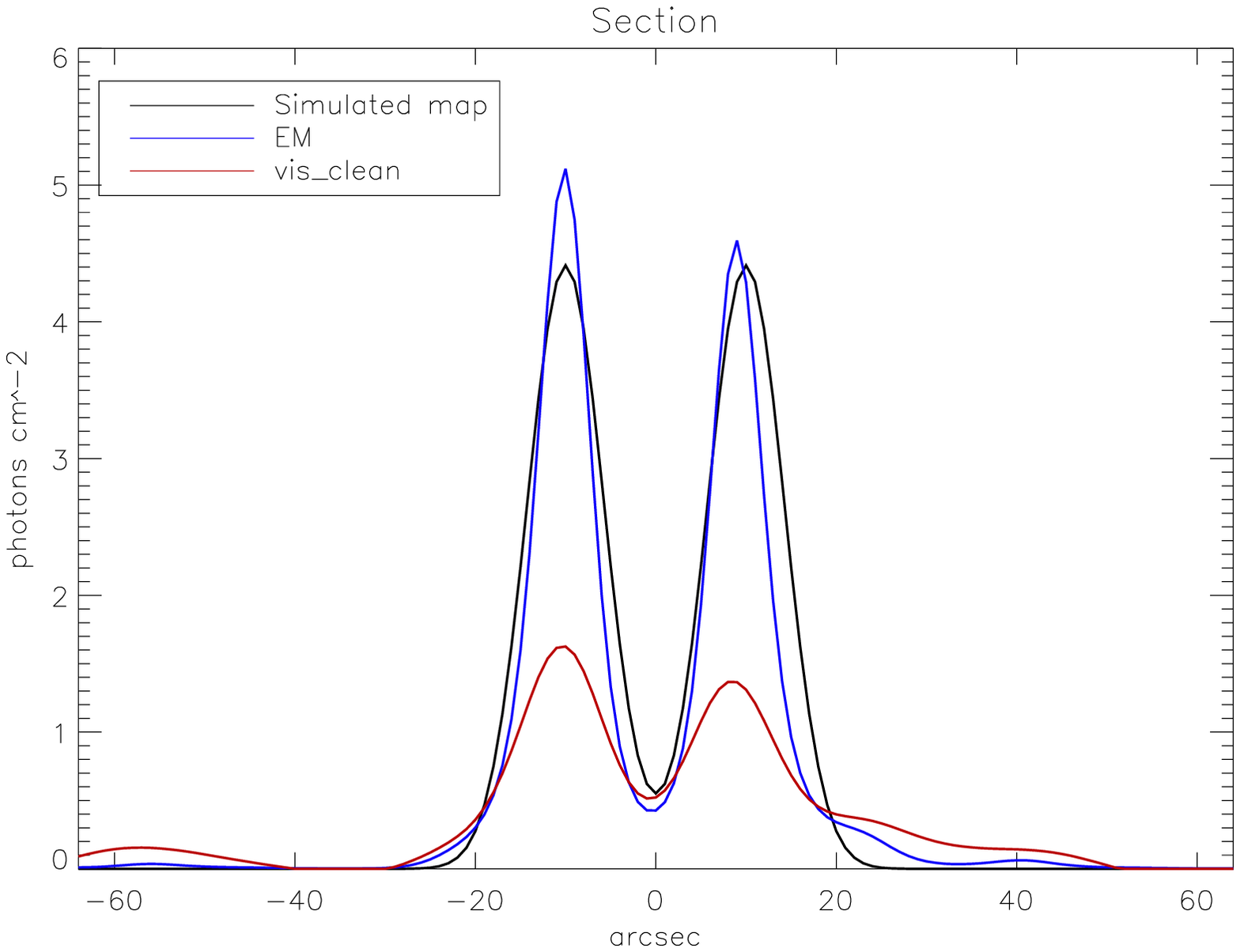} &
				\includegraphics[width=.3\textwidth]{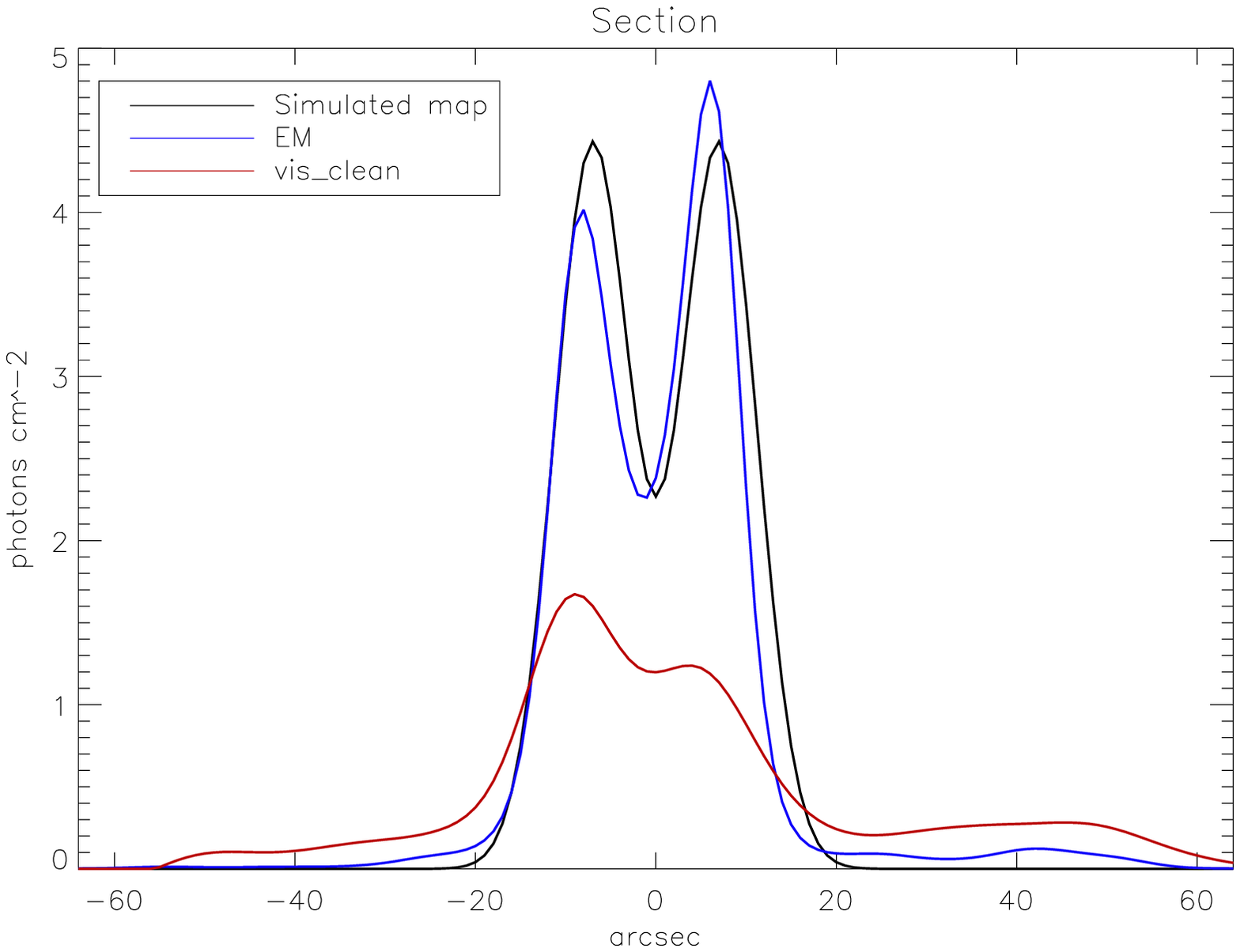} &
				\includegraphics[width=.3\textwidth]{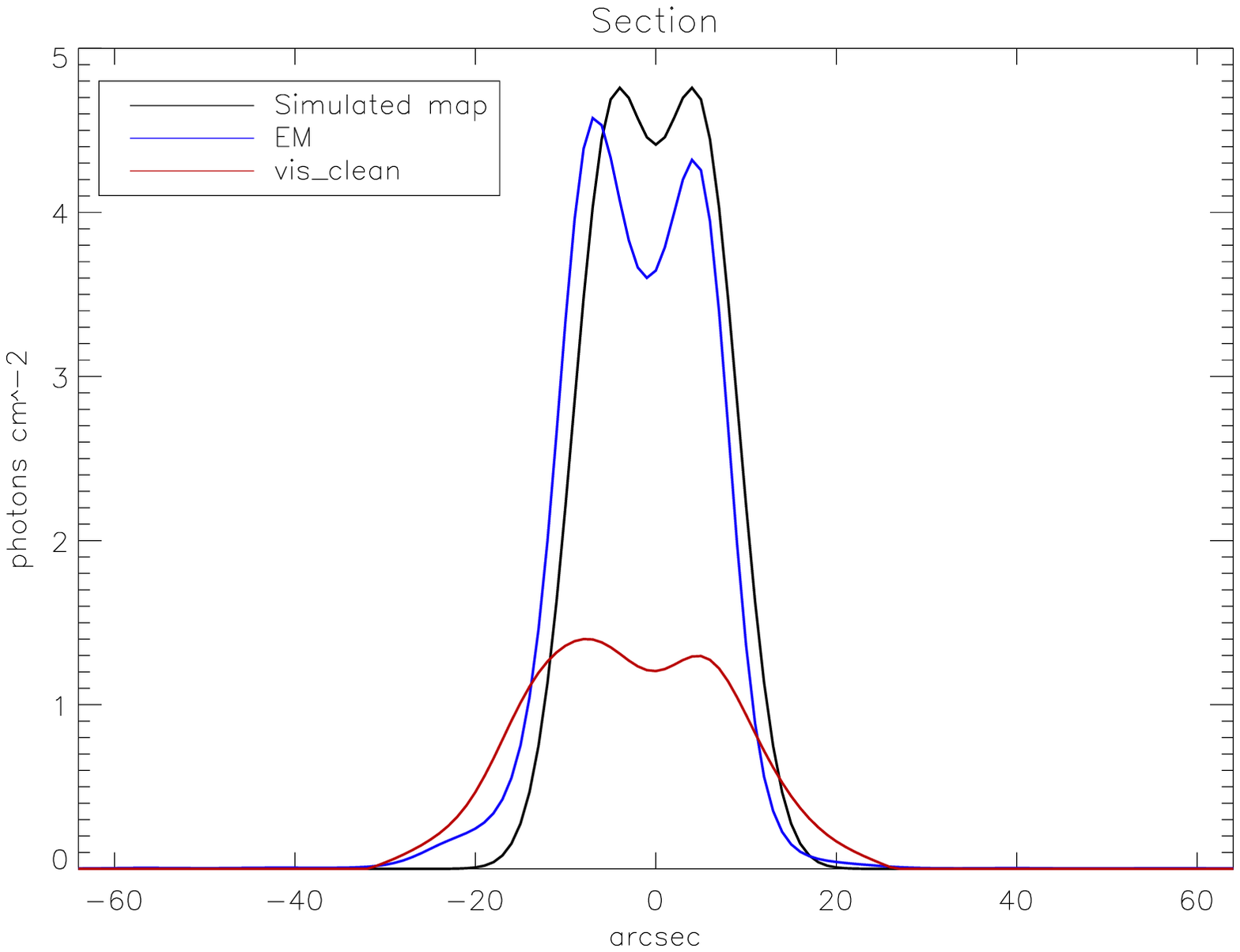}	
			\end{tabular}
			\end{center}
		\caption{Test on resolution power: two identical Gaussian functions with $FWHM = 10$ arcsec are moved closer and closer at $20$ arcsec (left column), $14$ arcsec (middle column), $10$ arcsec (right column), reconstructed by EM and CLEAN and compared to ground truth. The results concerning all three levels of statistic illustrate the reconstructed intensity profiles along the axis passing through the source centers, averaged over $10$ random data realizations. The corresponding confidence strips are also reproduced.}\label{fig:figure-3}
		\end{figure}

	\begin{figure}[h]
		\begin{center}
		\begin{tabular}{ccc}
			
				\includegraphics[width=.3\textwidth]{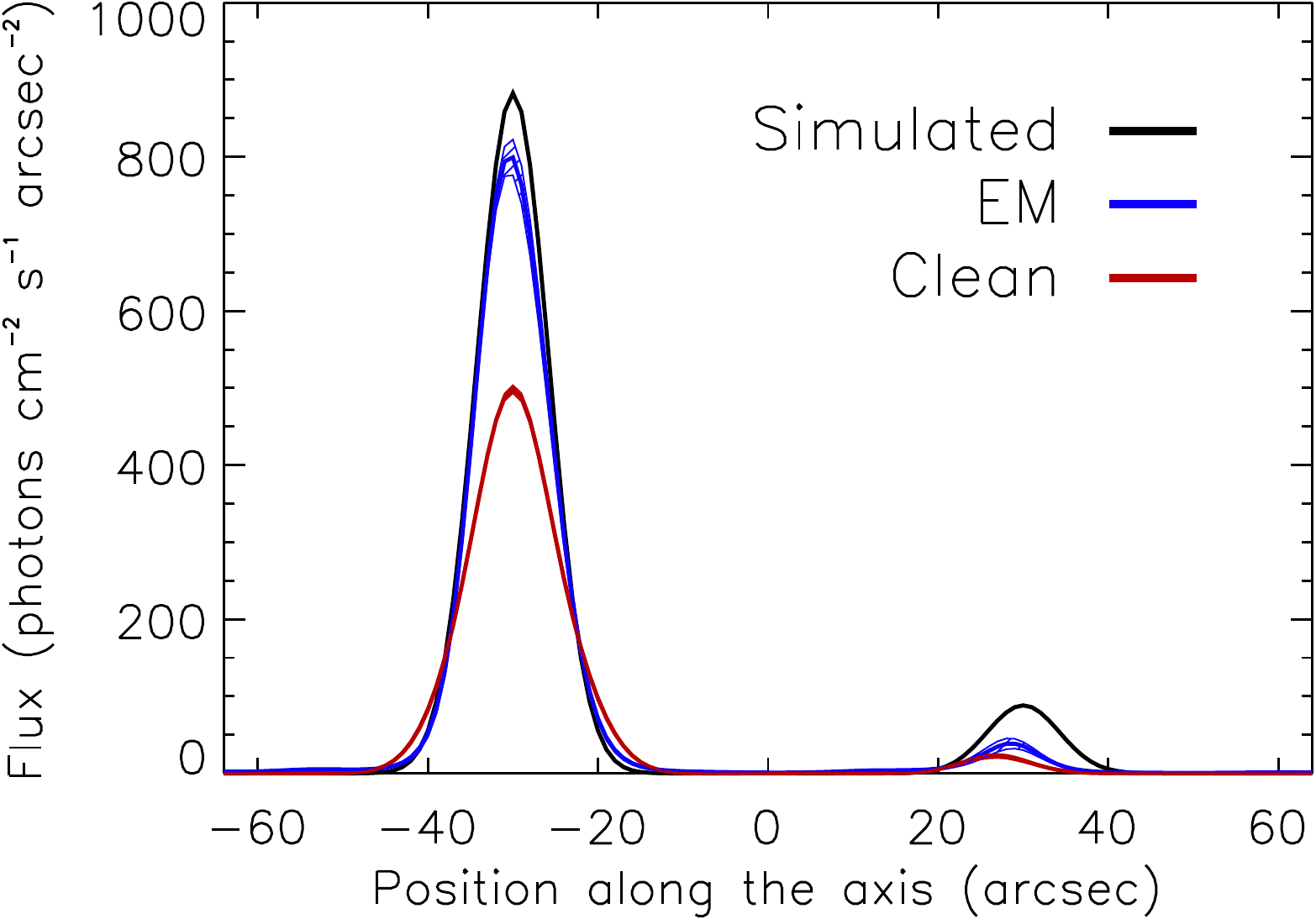} &
				\includegraphics[width=.3\textwidth]{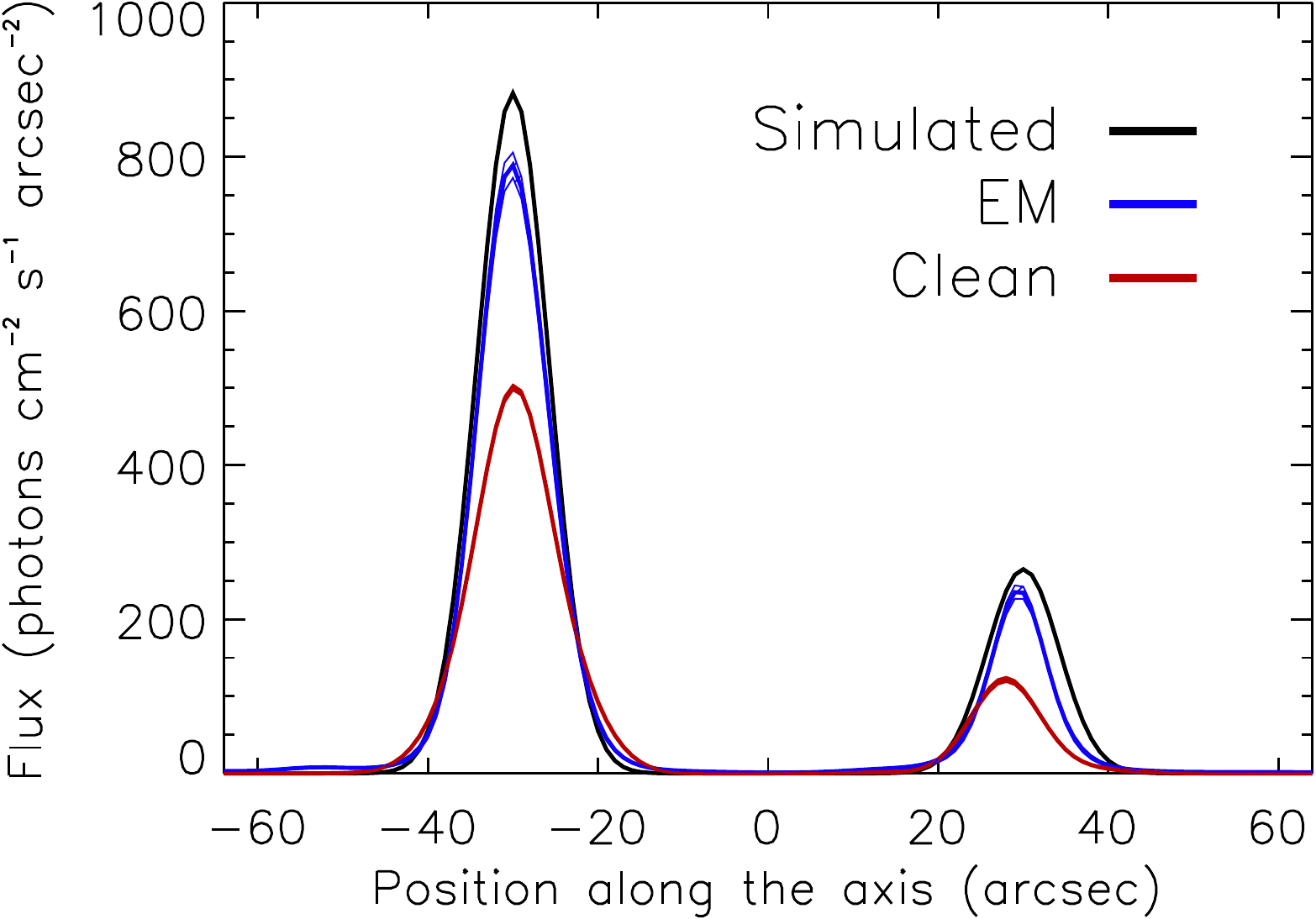} &
				\includegraphics[width=.3\textwidth]{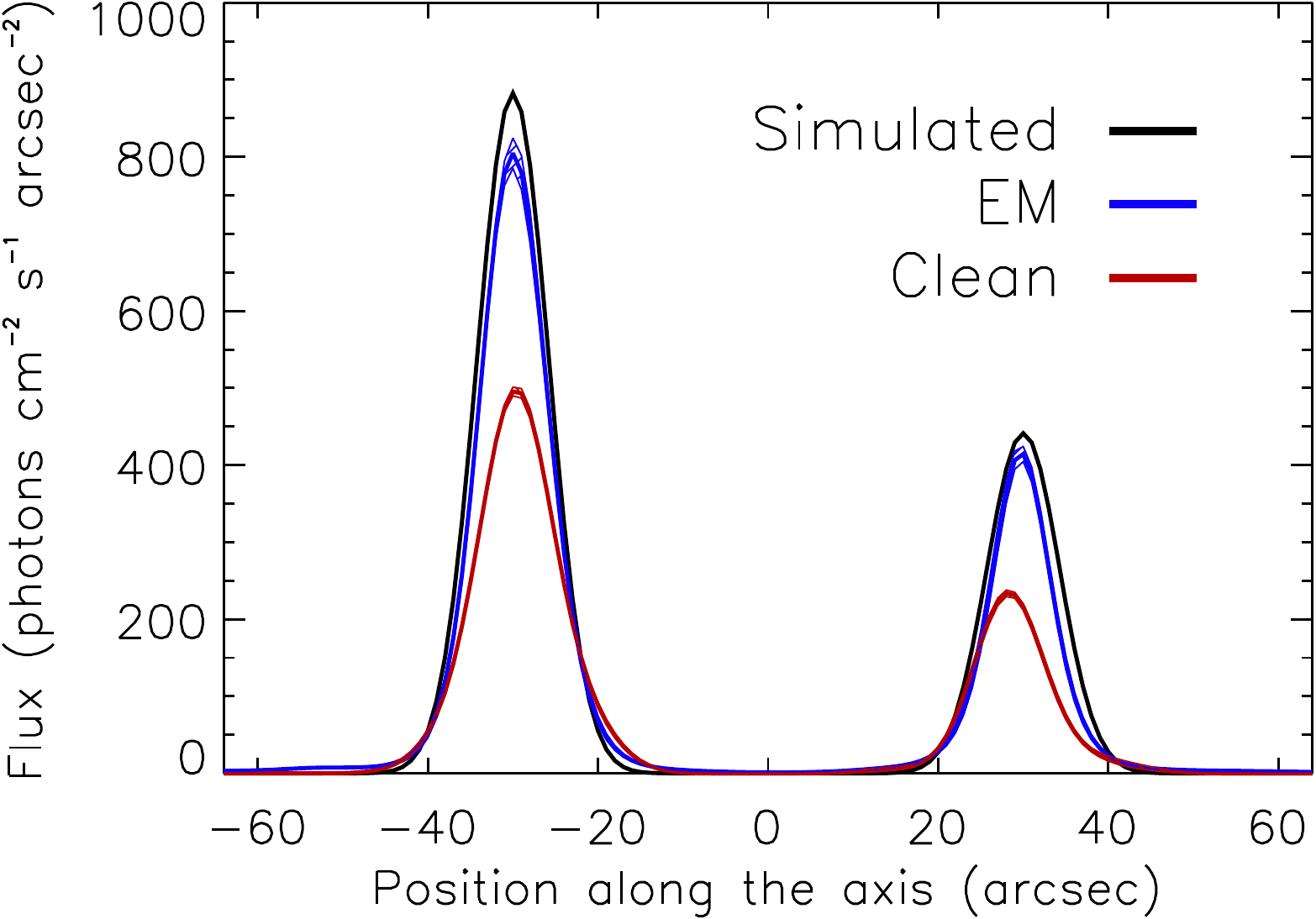}	 \\
			
			
				\includegraphics[width=.3\textwidth]{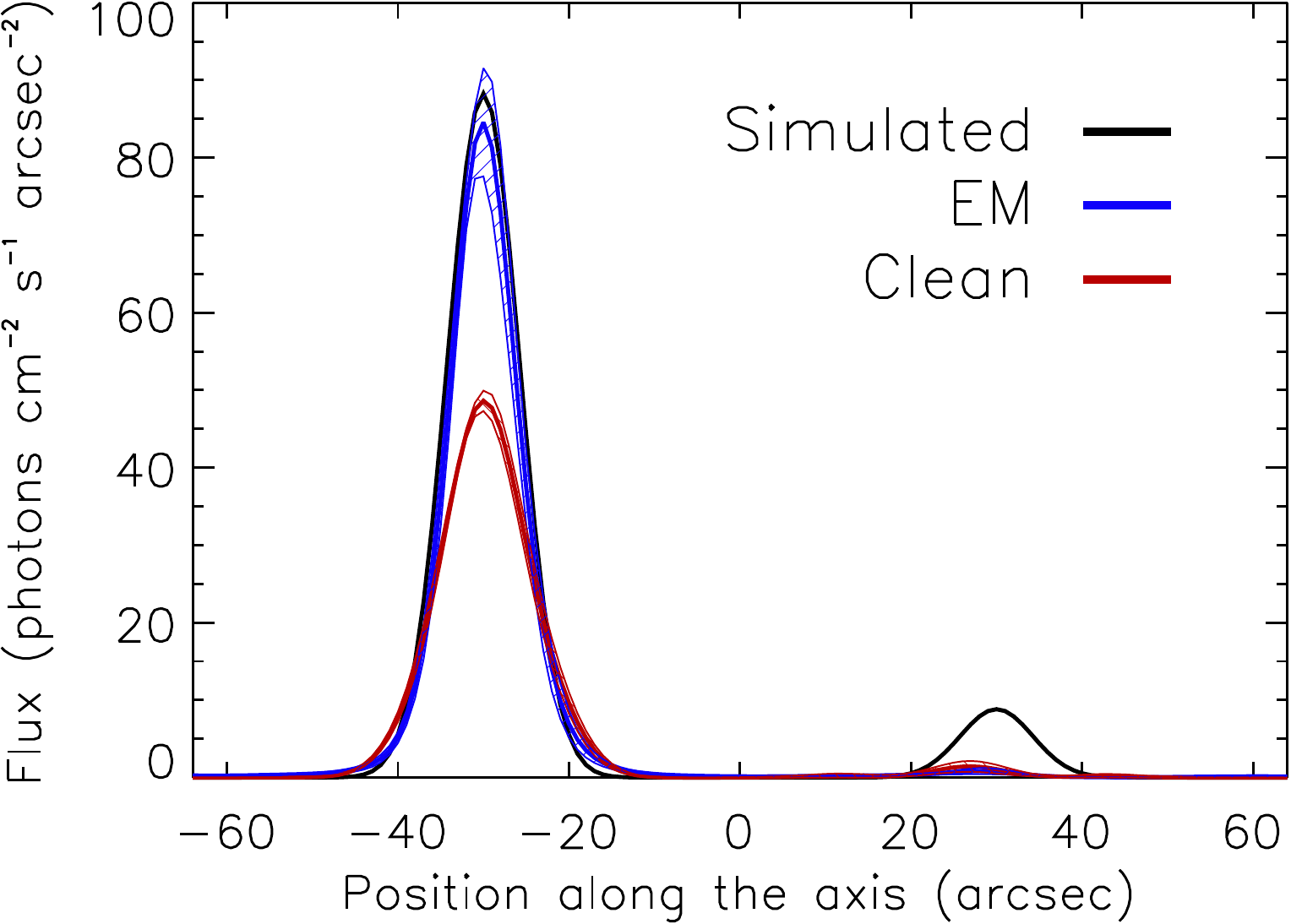} &
				\includegraphics[width=.3\textwidth]{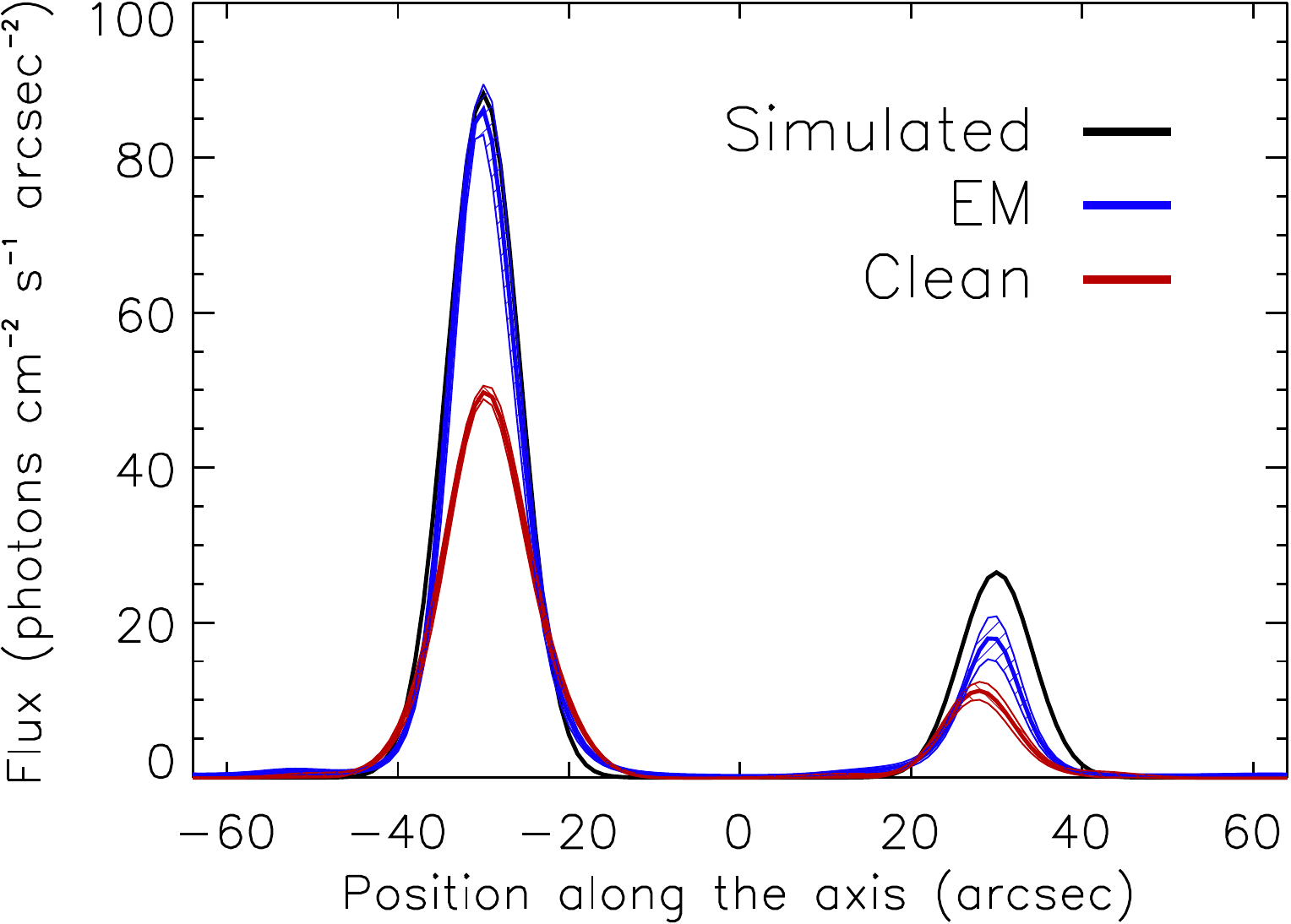} &
				\includegraphics[width=.3\textwidth]{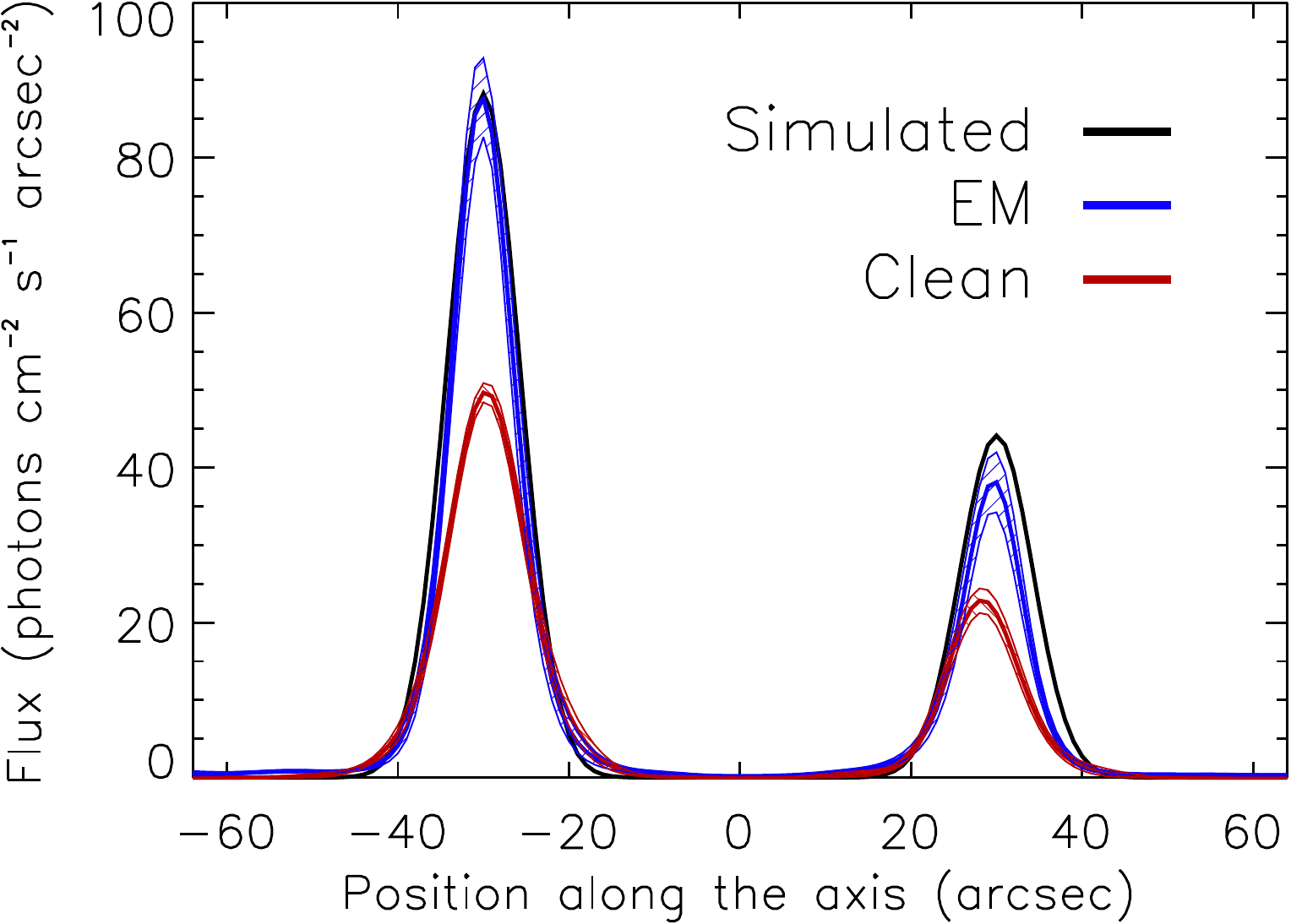} \\
			
			
				\includegraphics[width=.3\textwidth]{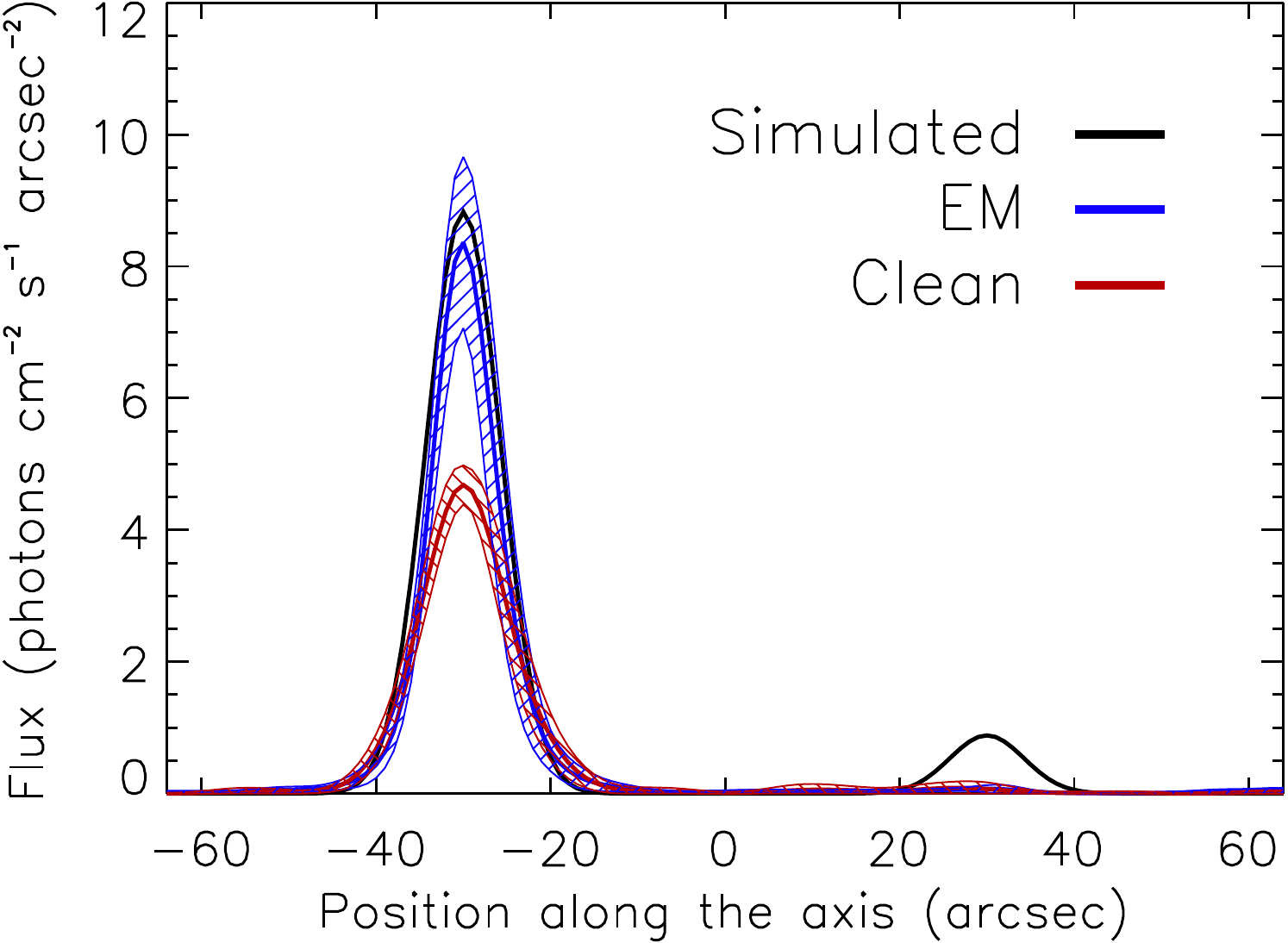} &
				\includegraphics[width=.3\textwidth]{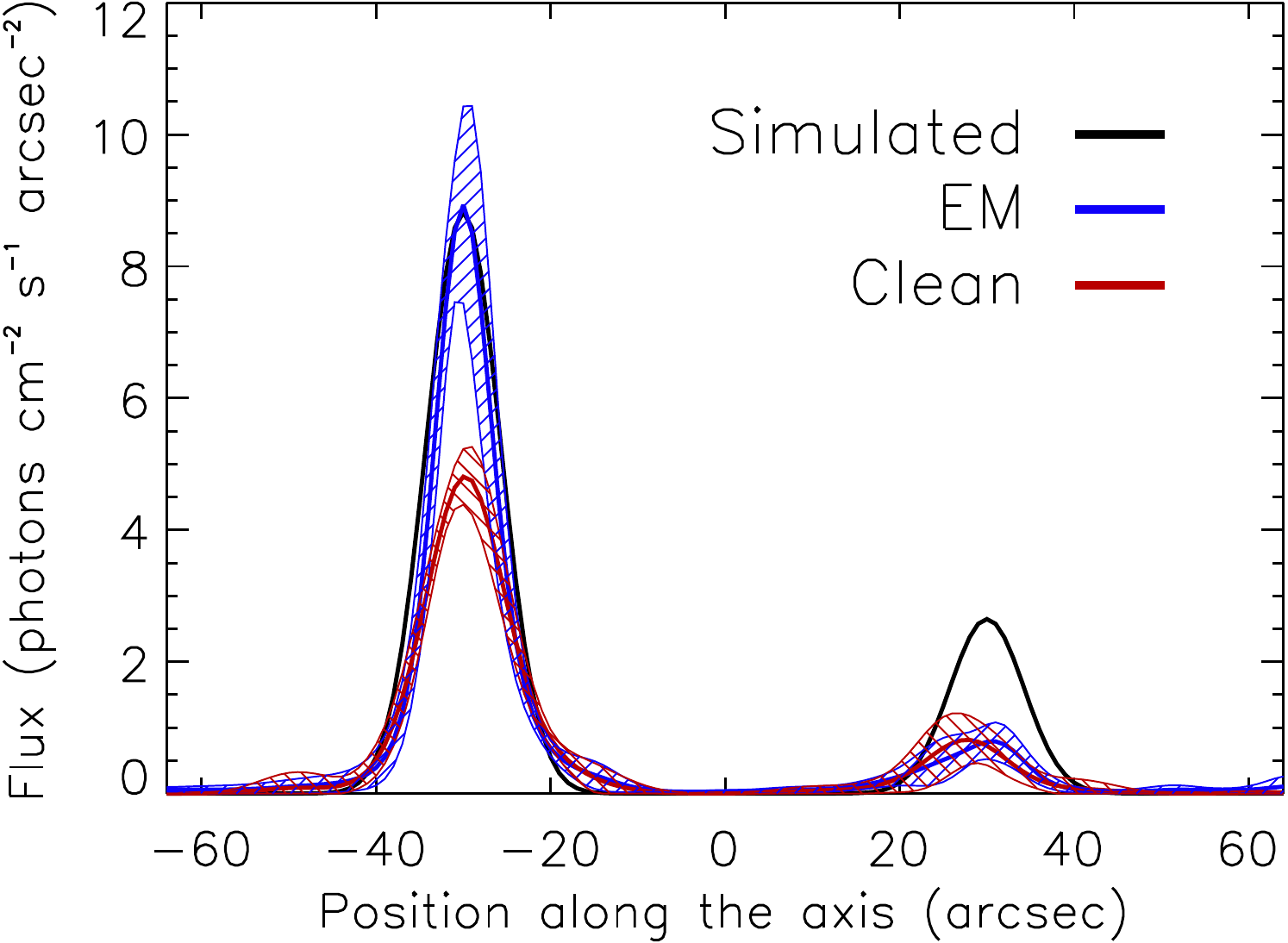} &
				\includegraphics[width=.3\textwidth]{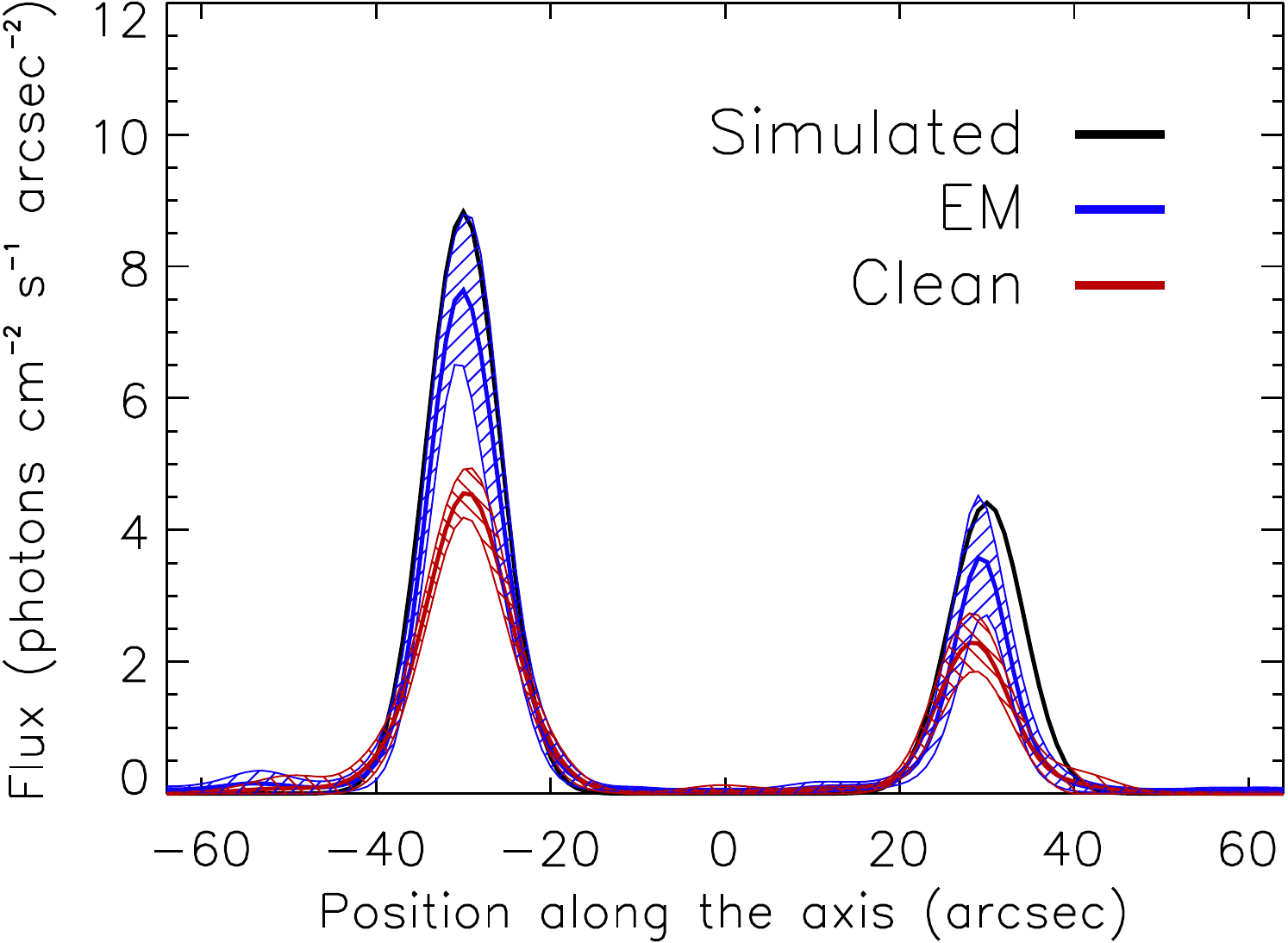}	
		\end{tabular}
		\end{center}	
		\caption{Test on dynamic range: two Gaussian functions with brightness ratio $b=0.1$ (left column), $b=0.3$ (middle column) and $b=0.5$ (right column) are reconstructed by EM and CLEAN and the results compared to ground truth. The results concerning all three levels of statistic illustrate the intensity profiles along the axis passing through the source centers, averaged over $10$ random data realizations.The corresponding confidence strips are also reproduced.}\label{fig:figure-4}
		\end{figure}

	
We tested EM on synthetic data simulated with the \textit{STIX} Data Processing Software (DPS). We considered four configurations: 
\begin{itemize}
\item a double foot-point flare (configuration FF1) in which the sources have the same flux but different size;
\item a double foot-point flare (configuration FF2) in which the sources have the same flux and the same size;
\item a loop-flare (configuration LF1) with large curvature;
\item a loop-flare (configuration LF2) with small curvature.
\end{itemize}

For each configuration we performed three simulations, corresponding to three different levels of the incident photon flux: high statistic refers to an overall incident photon flux of $10^5$ photons s$^{-1}$ cm$^{-2}$; medium statistic refers to an overall incident photon flux of $10^4$ photons s$^{-1}$ cm$^{-2}$; low statistic refers to an overall incident photon flux of $10^3$ photons s$^{-1}$ cm$^{-2}$. Further, for each configuration, we performed $10$ random realizations of the count data vector. In all cases we assessed the reliability of the reconstructions by evaluating the morphology, the photometry, the spatial resolution and the dynamic range provided by EM and comparing these results with both the ground truth and the results obtained by applying the most standard imaging algorithm implemented in the {\em{STIX}} DPS, i.e. CLEAN using visibilities as input. Figure \ref{fig:figure-2} shows the reconstructions for the four configurations when the input data correspond to the medium statistic level. The complete set of results can be reached in the 'Additional Materials' submitted to the Journal. In order to provide a more quantitative comparison with the ground truth and CLEAN reconstructions, Table \ref{table:table-1} contains the geometrical parameters associated to each original and reconstructed configuration together with the photometric parameters.

In order to assess the ability of the method to separate close sources, we simulated two identical circular Gaussian sources with $FWHM = 10$ arcsec whose peaks gradually approach from $20$ arcsec, through $14$ arcsec, to $10$ arcsec and made a spatial resolution analysis using EM and CLEAN. Results are presented in Figure \ref{fig:figure-3} for the three levels of statistic. In most conditions, the two methods reproduce rather well the locations of the sources but EM does systematically better than CLEAN in estimating the peak intensity. When the peak distance in the original sources is $10$ arcsec, EM is able to distinguish them at high statistic. 

Figure \ref{fig:figure-4} and Table \ref{table:table-2} illustrate the results concerning the assessment of the ability of the methods in reproducing the dynamic range of two competing Gaussian sources. We considered again two circular Gaussian sources with the same size ($FWHM = 10$ arcsec) but with brightness ratio (i.e., the ratio between the peak of the weakest source and the peak of the strongest source) $b=0.1, 0.3, 0.5$. Also in this case locations are reproduced rather well by both methods, and also in this case the peak intensities are better estimated by EM. As far as the estimation of the brightness ratio is concerned, neither method is reliable, at all statistics, when $b=0.1$; further, when $b=0.3$ and $b=0.5$ at at low and medium levels of statistic, CLEAN produces slightly better results.

\begin{table}[h]
	
	\begin{center}
		\begin{tabular}{ccccc}
			
			\toprule
			\textbf{statistic}		  &\textbf{algorithm}	&\multicolumn{3}{c}{\textbf{brightness ratio}} \\
			\cmidrule{3-5}				
			&						&\textbf{0.1} 		&\textbf{0.3}	 	&\textbf{0.5}\\
			\midrule
			\multirow{2}{*}{High} 	  &EM					&$0.049\pm 0.009$	&$0.30\pm 0.02$ 	&$0.52\pm 0.02$	\\	
			&Clean				&$0.045\pm 0.005$	&$0.24\pm 0.01$	&$0.47\pm 0.01$	\\
			\midrule
			\multirow{2}{*}{Medium}   &EM					&$0.010\pm 0.005$	&$0.21\pm 0.04$	&$0.44\pm 0.04$	\\	
			&Clean				&$0.03\pm 0.02$	&$0.23\pm 0.02$ 	&$0.46\pm 0.03$	\\
			\midrule
			\multirow{2}{*}{Low} 	  &EM					&$0.011\pm 0.007$	&$0.097\pm 0.028$ 	&$0.47\pm 0.10$	\\	
			&Clean				&$0.03\pm 0.03$	&$0.18\pm 0.08$ 	&$0.50\pm 0.09$	\\
			\bottomrule
			
		\end{tabular}	
	\end{center}
	\caption{Estimation of three brightness ratio values ($b=0.1,0.3,0.5$) provided by EM and CLEAN for three levels of statistic of the input photon flux. For each condition we performed the analysis in the case of $10$ random data realizations.}\label{table:table-2}
	
\end{table}

\section{Conclusions}
We presented a model of image formation for {\em{STIX}} in which the incoming photon flux distribution is mathematically projected onto the counts recorded by the detector pixels. This model presents the advantages of a better signal-to-noise ratio and of a higher number of input data at disposal for the reconstruction process with respect to the visibility-based model. Further, this approach allows the use of Expectation Maximization as imaging algorithm, and therefore the exploitation of the Poisson nature of the data statistic. The performances of EM show that the morphological parameters are reproduced with a level of detail comparable with the one provided by CLEAN. However, EM has a better photometric performance, in line with the fact that this method has been explicitly conceived for maintaining the overall count number during iterations. Interestingly, this good photometric behavior holds true even locally, as showed by the reconstructed values of the flux above $50\%$ level. Differently than CLEAN, EM simultaneously exploits both a positivity constraint and the conservation of flux at each iteration and this probably explains its fairly better performances in spatial resolution power: EM is able to  separate approaching sources in a rather nice fashion (note that in Figure \ref{fig:figure-3} the ground truth falls in the confidence strip of the EM reconstructions for most conditions). CLEAN systematically underestimates the peak intensity of the reconstructed sources. However, at medium and high levels of statistic for the incoming photon flux, CLEAN shows a slightly better performance in reproducing the brightness ratio.

In conclusion, the count-based imaging model for {\em{STIX}} provides new image reconstruction capabilities for an imaging instrument which has been originally designed for visibility-based approaches. Count-based methods can exploit the Poisson nature of data statistic in both Bayesian frameworks like EM and in deterministic settings that iteratively optimize the Kullbach-Leibler divergence. On the other hand, visibility-based methods are typically rather fast since they can exploit FFT. The development of further statistical and deterministic regularization techniques for the reduction of the count-based imaging model may be an interesting investigation theme for next steps of {\em{STIX}} imaging activity. 

\bibliographystyle{aa}
\bibliography{refs_aa}

%
%
%
%

\end{document}